\documentclass[12pt]{elsarticle}
\makeatletter
\def\ps@pprintTitle{%
 \let\@oddhead\@empty
 \let\@evenhead\@empty
 \def\@oddfoot{\centerline{\thepage}}%
 \def\@evenfoot{\thepage}
 \let\@evenfoot\@oddfoot}
\makeatother

\usepackage[final]{changes} 
\definecolor{C0}{HTML}{1F77B4} 
\definechangesauthor[color=C0]{R1}

\usepackage{hyperref}

\usepackage[T1]{fontenc}
\usepackage{lmodern}
\usepackage{microtype}
\usepackage[margin=1in]{geometry}
\usepackage{amsmath,amssymb,amsthm,bm}
\usepackage{graphicx}
\usepackage{subcaption}
\biboptions{sort&compress}
\usepackage{epstopdf}
\usepackage{float}

\AppendGraphicsExtensions{.tif}

\begin{document}

\title{Optimized and autonomous machine learning framework for characterizing pores, particles, grains and grain boundaries in microstructural images}
\author[auburn]{Roberto Perera}
\author[auburn]{Davide Guzzetti}
\author[auburn]{Vinamra Agrawal\corref{cor1}}
\cortext[cor1]{Corresponding author: vinagr@auburn.edu}
\address[auburn]{Department of Aerospace Engineering, Auburn University, Auburn, AL, USA}

\begin{abstract}
    Additively manufactured metals exhibit heterogeneous microstructure which dictates their material and failure properties. 
    Experimental microstructural characterization techniques generate a large amount of data that requires expensive computationally resources.
    In this work, an optimized machine learning (ML) framework is proposed to autonomously and efficiently characterize pores, particles, grains and grain boundaries (GBs) from a given microstructure image.
    First, using a classifier Convolutional Neural Network (CNN), defects such as pores, powder particles, or GBs were recognized from a given microstructure.
    Depending on the type of defect, two different processes were used.   
    For powder particles or pores, binary segmentations were generated using an optimized Convolutional Encoder-Decoder Network (CEDN).
    The binary segmentations were used to obtain particle and pore size and location using an object detection ML network (YOLOv5).
    For GBs, another optimized CEDN was developed to generate RGB segmentation images, which were used to obtain grain size distribution using two regression CNNs.
    To optimize the RGB CEDN, the Deep Emulator Network SEarch (DENSE) method which employs the Covariance Matrix Adaptation - Evolution Strategy (CMA-ES) was implemented. 
    The optimized RGB segmentation network showed a substantial reduction in training time and GPU usage compared to the unoptimized network, while maintaining high accuracy.
    Lastly, the proposed framework showed a significant improvement in analysis time when compared to conventional methods.
\end{abstract}


\maketitle

\section{Introduction} \label{Introduction}
    Over the past decade, additive manufacturing of metallic materials has found applications towards aerospace rocket components, automotive parts, biomedical equipment, and infrastructural components \cite{Russell2019Qualification, Gasser2010Laser, Su2012Development, Woensel2018Printing}.
    The resulting microstructure of additively manufactured (AM) components, which governs its mechanical properties, is intrinsically linked to the manufacturing process itself. 
    For instance, the size distribution and morphology of powder particles plays a significant role in mechanical properties of the components manufactured using powder bed fusion (PBF).
    These powder particles not only cause alterations in thermal conductivity and energy absorption, but also influence porosity, surface roughness, hardness, and strength of the final built part \cite{DeCost2017Characterizing, Murr2018Metallographic}. 
    Additionally, imperfections in the manufacturing process such as trapped inert gases during solidification may result in higher porosity.
    Microstructural pores are known to decrease material properties such as Young's Modulus, strength, and fatigue life \cite{Choren2013Youngs, FURUMOTO2015Permeability, Tammas-Williams2017FatiguePorosity}.
    Pores also introduce stress concentrations that may lead to crack nucleation \cite{SHERIDAN2018Fatigue}.   
    Moreover, the build direction utilized during manufacturing results in the formation of columnar grain structures \cite{Todaro2020Grain}.
    These grain structures are characterized by microstructural grain boundaries (GBs), which influence material properties such as hardness, fatigue strength, and stress-strain relations \cite{Chapetti2004Fatigue}.
    Therefore, characterizing microstructural features in AM metallic components is critical to determining its mechanical properties and performance.
    
    A vast number of imaging techniques have been developed in the past to characterize these microstructural features. 
    These include Helium pycnometry, scanning electron microscopy, energy dispersive spectroscopy, electron back scattering diffraction, laser diffraction, X-Ray computed tomography, and X-Ray diffraction \cite{Slotwinski2014Tools}.
    These techniques have proved to be extremely useful in the microstructural characterization of AM materials.
    However, they generate a large amount of data, which requires expensive computational time for post-processing.
    Image segmentation is a valuable computational technique utilized for the characterization of pores, particles and GBs.
    Image segmentation provides analysts with rapid results of the material's microstructure through visualization techniques.
    Well known algorithms such as city-block distance function + watershed segmentation \cite{RABBANI2014An} and assisted threshold binarization procedure \cite{LORE2015Facile} employ image segmentation to identify pores.
    For GB identification, the point-sampled intercept length method (PSILM) \cite{Gundersen1985Stereological, ASTM2015Standard} is most widely used. 
    This algorithm first uses a standard edge detection algorithm (for instance, a standard MATLAB built-in tool) to place a fixed number of points overlapping the detected edges.
    For each pair of points, it then generates lines along four evenly spaced directions ($0^{\circ}$, $90^{\circ}$, $45^{\circ}$, $-45^{\circ}$).
    Finally, it uses the the resultant intersecting lines to generate histograms for the grain size and their RGB segmentations. 
    
    Although conventional segmentation tools have proved their use in algorithms for feature characterization of pores, particles, and grain boundaries, they are restricted to time-consuming pre-processing image operations such as gamma corrections, edge detection, threshold value selection, interpolation functions, morphological operators, and others \cite{RABBANI2014An, LORE2015Facile, Gundersen1985Stereological, ASTM2015Standard}.
    Lastly, an important disadvantage from these methods is that the user must know prior to each analysis the specific microstructural features (pores, particles, or grains) to be extracted, to then apply the corresponding conventional method.
    Therefore, it is crucial to develop a framework that may autonomously characterize microstructures regardless of their defects, while improving analysis time and reducing computational expenses. 
    
    ML offers a way to dramatically improve data processing efficiency \cite{Qiu2016BigData} and shorten the turnaround time, while achieving accuracy comparable to conventional methods without prior information input from the operator.
    In the past, various ML methods have been implemented for microstructural feature characterization.
    Examples include classification of AM powder feedstocks with respect to their particle size distribution (PSD) \cite{DeCost2016Feedstocks}, classification of microstructural dendrites (vertical, horizontal or none) \cite{CHOWDHURY2016ImageDriven}, and classification of coarse-grained dislocation microstructures in terms of different dislocation density field variables \cite{Steinberger2019Dislocations}.     
    However, these attempts were limited in their ability to extract more than one detail of the microstructure.
    For instance, the spatial distribution and clustering of grains influences the material's hardness \cite{Furukawa1996Microhardness,Tachibana1988Effects} which is not captured by these methods.
    
    {
        To enable detection of multiple microstructural features, ML techniques have been integrated with conventional image processing tools.
        An example of this approach was demonstrated by Wei Li, Kevin G. Field, and Dane Morgan \cite{Li2018Automated}.
        They showed that by integrating a cascade detector, a 15-layer CNN, a watershed flood algorithm, and the MATLAB built-in tool regionprops, the extraction of shape, size, and location of loops and line dislocations in irradiated steel was possible.
    }
    {   
        Another similar example, is the development of the automated detection workflow for helium bubbles in Irradiated X-750 \cite{ANDERSON2020Automated}. 
        In this work, the Region Proposal Faster R-CNN network was used along with the open source program LabelImg to generate bounding boxes for each defect with orders of magnitude faster than manual analysis. 
    }
    
    ML encoder-decoder networks and CNNs have also been used in the past for binary or RGB segmentations for microstructural characterization.
    For example, the U-Net encoder-decoder network was used to obtain binary segmentations of aluminum alloys \cite{Li2020supportvector}.
    Fully Convolutional Neural Networks (FCNNs), and Residual Neural Networks (RNNs) have also been implemented to characterize the shape of microstructures as "lamellar”, “duplex”, or “acicular”, to then obtain binary segmentations \cite{BASKARAN2020Adaptive,Jang2020Residual}. 
    {
        Another example is DefectSegNet \cite{Roberts2019DefectSegNet} which used a CEDN to generate binary segmentations of line dislocations, and precipitates or voids in SEM images of steel. 
        This network, which involved the combination of the state-of-the-art networks, U-Net and DenseNet, outperformed human expert defect quantification with faster performance. 
    }
    
    \replaced[id=R1,comment={Q1c}]{However, }{Some of}  the main challenges of these current ML segmentation methods are their high GPU usage, prolonged training time, and architecture's complexity.
    For example, the U-Net requires a training time of  1:55 hours, and GPU usage of 998.28 MB, in order to apply RGB segmentations of grain boundary microstructures (Section \ref{Subsec:4.1} and \ref{Subsec:4.3}).
    {While the development of new NVIDIA GPU models in the recent years has improved GPU speed and memory requirements, as problems scale to more complex networks and larger datasets, the need for faster and simpler networks may be a benefit.
    Free GPU resources such as Google Colab provide up to 12 GBs of GPU memory for 12 hours, and 24 hours with the purchase of the professional account. 
    Developing more computationally efficient networks that perform with similar accuracy as current state-of-the-art architectures may benefit applications running on resource-bound machines or pay-by-use contracts.
    Similarly, similar networks for the same performance may yield a smaller resource utilization footprint if deployed to a device.
    }
    
    In 2020, M. F. Kasim demonstrated a rapid deep learning environment called Deep Emulator Network SEarch (DENSE) \cite{Kasim2020Building} that offered potential to reduce training time and GPU usage by orders of magnitude.
    In this work, we leverage DENSE to reduce training time and GPU usage of the current state-of-the-art segmentation models and encoder-decoder networks.
    We obtain simpler architectures that may provide with high-accuracy RGB segmentations for GBs, and binary segmentations for particles and pores, while substantially decreasing analysis time.
    We use the standard PSILM algorithm as a baseline to generate a large dataset of GB features \cite{Lehto2014Influence, Lehto2016Characterization}.
    We use the conventional city-block distance function in combination with the watershed segmentation method \cite{RABBANI2014An} as a baseline to generate a large dataset for particles' or pores' features.
    Next, we use ML classifier CNNs to recognize the type of defects present in the given microstructure such as pores, particles, or grain boundaries.  
    We simplify the conventional segmentation encoder-decoder network U-Net for binary segmentations of pores and particles to reduce the network's complexity, GPU usage, and training time.
    Leveraging DENSE, we optimize the simplified encoder-decoder network for RGB segmentations of GBs.
    Finally, by combining these ML tools along with regression CNNs, we develop a framework for heterogeneous microstructural feature characterization of multiple types of defects without prior user-defined input. 
    
    The paper is organized as follows. 
    We explain the ML networks and optimization methods implemented in the framework in Section \ref{Sec:2}. 
    We present an overview of the framework's structure and the process of gathering the training data in Section \ref{Sec:3}.
    We then present the results of each ML network's training parameters, GPU usage, training time, and accuracy analysis in Section \ref{Sec:4}.
    Finally, we present the framework's performance on multiple materials containing different types of defects, as well as a time-consumption analysis when comparing the framework's performance to the conventional standard PSILM in Section \ref{Sec:4}.

\section{ML methods and optimization} \label{Sec:2}

    \subsection{Classifier CNN, and Regression CNNs} \label{Subsec:2.1}
        Classification CNNs are one of the most popular deep supervised learning techniques in ML, with applications in 
        handwriting recognition, spam emails, fashion clothing, person's gender, to even real-time object recognition \cite{Al-Wzwazy2016Handwritten,Shang2016Spam,Bhatnagar2017Fashion,Levi2015Age,Sharma2018RealTime}.   
        This method can not only predict a binary output, but also up to thousands of classes with the use of the ``SoftMax'' activation function \cite{Kaibo2003SoftMax}.
        Another popular deep supervised learning technique in ML is the implementation of regression CNNs.
        While classification focuses on the prediction of integers, regression CNNs focus on the prediction of a decimal number.
        Some of the most famous applications of regression CNNs include the detection of facial landmarks, head-pose or human-pose landmarks, and house stock market prediction \cite{Xia2019LandmarkPose, Chenjing2014AgeCNN, Chen2018StockMarket}.
        
        We use a classifier CNN to classify a given microstructure feature as pores, particles, or GBs. 
        This feature may enable analysts to use a mixed dataset of multiple defects without the need for prior separation or classification of the microstructures.
        We then use two regression CNNs to generate histograms of radius values (x-axis), and the frequencies (y-axis) for the grain boundaries' distribution.
        
    \subsection{ML Encoder-Decoder Networks: U-Net and Res-UNet} \label{Subsec:2.2}
        ML Encoder-Decoder networks for semantic segmentation of images were introduced in 2015 with the development of the SegNet \cite{Cipolla2017SegNet}.
        The SegNet method was motivated by applications of scene understanding using real-time ML image segmentation of objects, animals, people, and nature.
        This new method gave rise to several encoder-decoder networks aimed at image segmentation for specific tasks.
        One of the most famous architectures following the SegNet was the U-Net; also introduced in 2015 by adding fully connected convolutional layers and feedforward layers between the encoder-decoder section \cite{Ronneberger2015unet}.
        While the SegNet focused on scene understanding of real-time images, the U-Net was developed specifically for biomedical segmentation of microscopic neuronal structures.
        
        Over the years, U-Net has been augmented with other ML techniques to improve its performance on specific tasks.
        Residual Networks (ResNet) is a popular example of the augmented techniques that 
        replaced the fully connected, and feedforward layers following the encoder section \cite{He2015Residual}
        to avoid vanishing or exploding gradients using skip connectors and identity layers \cite{Philipp2017Gradients}.
        This method known as the Res-UNet, has proved to be useful in complex biomedical segmentation applications such as retina vessels, photoacoustic and photovoltaic panels, and brain tumors \cite{Xiao2018Retina, Feng2020Photoacoustic, Zhang2019Photovoltaic,Hai2019BrainTumor}.
        
        In the proposed framework, we will optimize the conventional U-Net and Res-UNet networks for the tasks of binary and RGB image segmentation of microstructural pores or particles, and grain boundaries. 
            
    \subsection{Deep Emulator Network SEarch (DENSE)} \label{Subsec:2.3}
        Deep Emulator Network SEarch (DENSE) was introduced in January 2020, as an optimization method for deep neural networks' architectures aimed to emulate large-scale system dynamics \cite{Kasim2020Building}. 
        The idea was to speed up current computationally expensive simulation models for astrophysics, climate change, biogeochemistry, high energy density physics, and seismology, by searching for the optimal network's architecture parameters.
        While ML has recently showed its potential to speed up computer simulation models, optimization of the network's architecture parameters such as the number of channels, dimension of filters, size of padding, and size of stride was a challenge due to the time required to process each simulation.
        The DENSE method has been shown to speed up this task by introducing the probabilistic evolution algorithm of Covariance Matrix Adaptation - Evolution Strategy (CMA-ES) to rapidly generate a search space of network architecture parameters \cite{Hansen2016CMA}.
        We will use the DENSE algorithm in this framework to optimize the binary and RGB encoder-decoder segmentation networks, and the regression CNNs.    
        
    \subsection{Simple - UNet, and DENSE - UNet} \label{Subsec:2.4}
\begin{figure}
            \centering
            \includegraphics[width=0.7\linewidth]{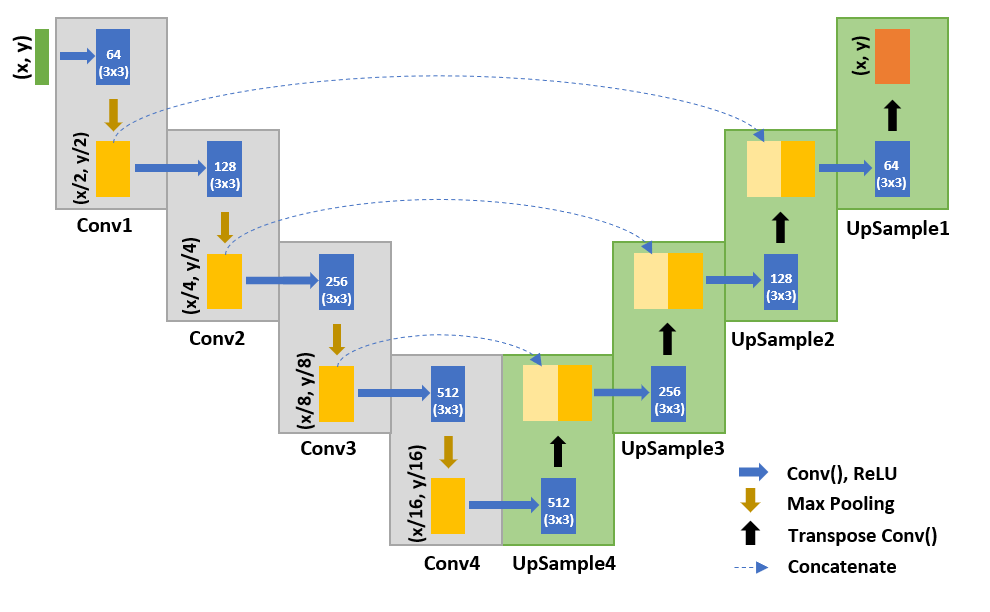}
            \centering
            \caption{Network architecture of the simplified U-Net encoder-decoder, Simple-UNet.}
            \label{fig:Simple-UNet}
        \end{figure}

        We developed the encoder-decoder network Simple U-Net to obtain binary segmentations for pores or particles.
        We simplified the network's architecture from the conventional U-Net by removing 4 convolution layers, 4 transpose convolution layers, and the fully connected feedforward layers as shown in Figure \ref{fig:Simple-UNet}. 
        This reduced the U-Net's complexity, GPU usage, and training time while effectively applying binary segmentation.
        Thus, making the Simple-UNet, the prime encoder-decoder candidate for the segmentation of pores or particles.
        
        Finally, we use the DENSE algorithm to optimize the Simple-UNet's architecture extending its capabilities to RGB segmentation of GBs.
        We optimized the architecture with respect to the segmentation accuracy for GBs, by varying the network's filters using (1x1), (3x3), (5x5), (7x7), and (9x9) dimensions within the DENSE algorithm's parameters.
        Additionally, we kept the network's channels the same as for the Simple-UNet shown in Figure \ref{fig:Simple-UNet}. 
        
    \subsection{YOLOv5} \label{Subsec:2.5}
        Region-based Convolutional Neural Network (R-CNN) opened the door to object detection ML algorithms \cite{Girshick2014Rich}.
        Shortly after, the YOLO (You Only Look Once) algorithm was developed to speed up the object detection process.
        YOLO introduces a single regression CNN pass to approximate the best size for the detection box \cite{Redmon2015You}.
        Over the years, YOLO algorithm has been improved in various aspects: number of objects (YOLO9000), prediction accuracy (YOLOv3), speed of real-time analysis (YOLOv4), and memory requirements and exportability (YOLOv5)
        \cite{Redmon2016YOLO9000, Redmon2018YOLOv3, Bochkovskiy2020YOLOv4, Jocher2020yolov5}.
        We leveraged YOLOv5's GPU memory efficiency to detect pores or particles from binary segmentation.

    \subsection{Google Colab and Pytorch} \label{Subsec:2.6}
        Lastly, Google Colab, short for ``Collaboration'' is a free online programming service provided by Google Research and hosted by Jupyter notebooks.  
        It proves to be a useful ML learning tool not only for beginners, but also for advanced projects by providing 12GB of NVIDIA Tesla k80 free GPU for up to 12 hours \cite{Bisong2019Google}.
        %
        Pytorch \cite{NEURIPS2019PyTorch} is one of the most popular recent ML libraries for research applications as it is expressed in idiomatic Python programming language using dynamic computation graphs.
        Pytorch also defines a new class of "tensors", similar to "NumPy" arrays, with the addition of CUDA-capable Nvidia GPU operations to speed up training process \cite{Paszke2017Automatic}.
        %
        We used Google Colab along with the Pytorch library for the development, analysis, and optimization of the proposed ML framework.

\section{Microstructural feature characterization} \label{Sec:3}

    \subsection{Training data and augmentation process} \label{Subsec:3.1}
        We used two separate datasets for powder particles and GBs to train each of the ML networks involved in the framework.
        For powder particles, we used a set of synthetic SEM material images \cite{Holm2016Synthetic} 
        containing 2048 synthetic microstructural images, separated into 8 categories of varying PSD (available online through Mendeley Data).
        We used the watershed and city-block distance (WCBD) segmentation method \cite{RABBANI2014An} to generate binary segmentations to train the Simple-UNet, and bounding box parameters to train YOLOv5 network.
        We modified the existing algorithm for the segmentation of pores by inverting the output to black pixels for the defects' location, and white pixels for the surrounding space.
        We also extracted and saved arrays containing the centroid coordinates, and respective diameters of each pore to train the YOLOv5 network.
        Lastly, we used the free online ``Roboflow.ai'' tool to generate the dataset of bounding boxes in the format compliant with the YOLOv5 network. 
        
        For GBs, we used the dataset \cite{Li2018Selective} containing 59 high resolution 2048x1532 images of selective laser melting (SLM) 316L stainless steel (available online via Mendeley Data).
        We then resized 
        each image to 2048x2048, cropping each section into separate 256x256 images, to obtain a total of 3776 images. 
        Finally, we used the standard PSILM algorithm \cite{Lehto2014Influence, Lehto2016Characterization} to obtain the ground truth RGB segmented images and the histograms for their radius size distribution.  
       
        {
            To train the Classifier CNN, 2000 randomly chosen images of both powder particles, and grain boundaries, were combined to obtain a total of 4000 images. 
            The training, validation, and test sets were evenly distributed by 3800, 100, and 100, respectively.
            To train the Simple-UNet for binary segmentations, each of the 512x512 synthetic particle images were cropped to 256x256, resulting in a total of 8192 images.
            This dataset was then distributed by 7850 for the training set, 512 for the validation set, and 100 for the test set.
            Additionally, since the YOLOv5 allows the use of smaller datasets to achieve high accuracies, the training, validation, and test sets were distributed as 800, 100, and 100 images, respectively. 
            
            Similarly, to train the DENSE-UNet for RGB segmentations, and the Regression CNNs for prediction of grain size distribution, the training, validation, and test sets were distributed as 3560, 108, and 108 images, respectively.
        }        

    \subsection{Overview of microstructural feature characterization framework} \label{Subsec:3.2}
\begin{figure}
            \centering
            \includegraphics[width=0.98\linewidth]{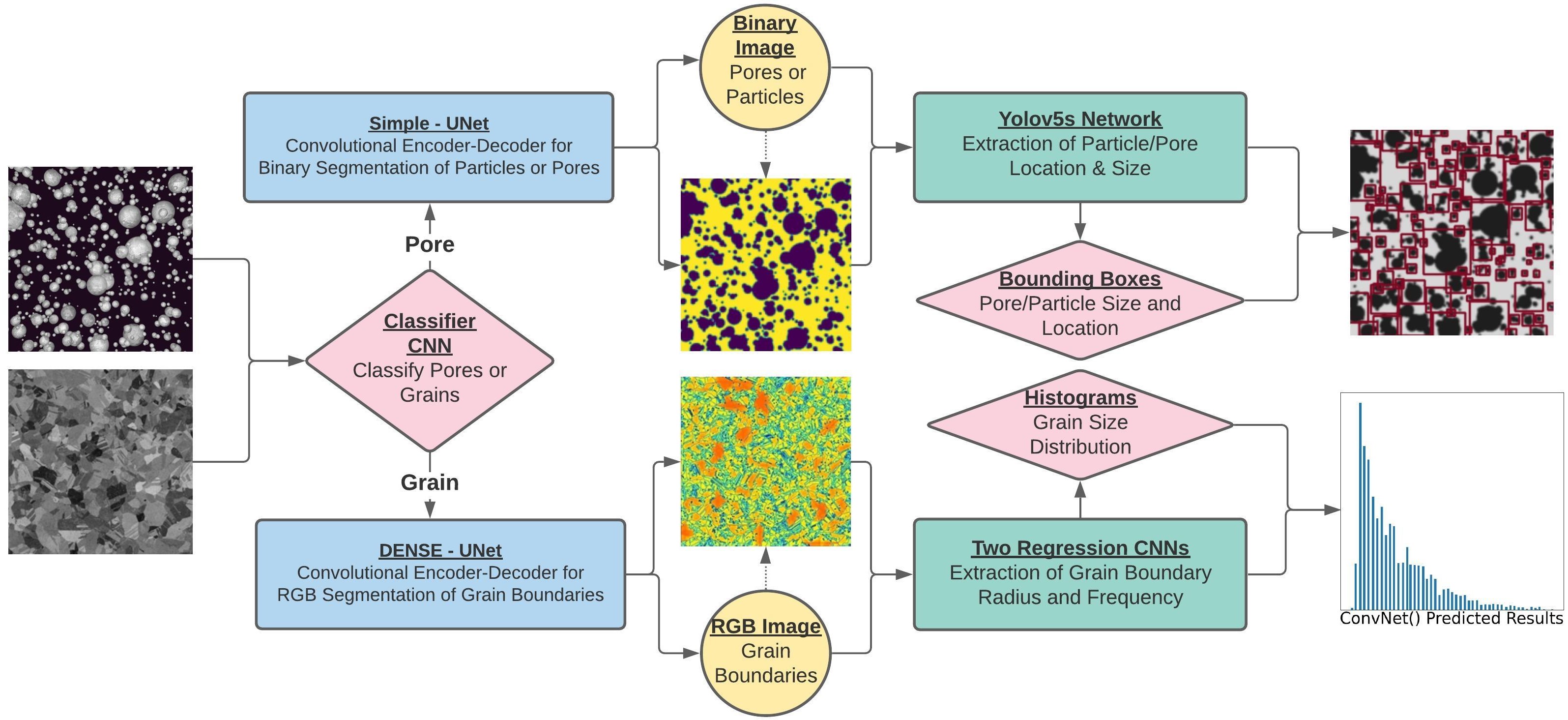}
            \centering
            \caption{Flowchart of the machine learning framework structure for microstructural feature characterization}
            \label{fig:framework}
        \end{figure}

        The developed framework follows the structure shown in Figure \ref{fig:framework}.
        The first advantage of the framework, is that no prior knowledge or specifications for the microstructural features in the images are required.
        We achieved this by integrating a CNN as the first ML network in the framework, labeled as ``Classifier CNN''.
        The classifier CNN processes a given microstructural image and categorizes it as pores, particles, or grains.
        Once the image has been correctly classified, we outputted it to two separate analysis paths. 
        
        For pores or particles, the path involves the simplified encoder-decoder network (Simple - UNet) to obtain binary segmentation. 
        As shown in Figure \ref{fig:framework}, the segmentation labels the locations that include particles as black pixels, and the void space surrounding the particles as white pixels. 
        Next, we used the YOLOv5 network 
        to extract the specific centroid location of each single pore, or of overlapping pores, as well as their size as shown in Figure \ref{fig:framework}.
        
        The second path characterizes the grains and GBs by performing RBG segmentation using the optimized encoder-decoder network, DENSE - UNet.
        Unlike pores, where the YOLOv5 algorithm was sufficient to extract their size and location, a more complex approach was necessary to characterize grains and GBs.
        This was due to the fact that particles, or pores are usually dispersed throughout the material's microstructure allowing the use of separated bounding boxes surrounding each feature.
        However, GBs are usually continuous throughout the material microstructure.
        We used the RGB segmentations of GB images to extract their distribution and clustering throughout the material's microstructure from the resulting color variations. 
        As shown in Figure \ref{fig:framework}, the color ranges from dark blue for smaller grains, to dark red for larger sized grains. 
        Finally, we used two regression CNNs to predict the radii (microns) and their frequencies to generate grain size distribution histograms from the resultant RGB segmentations.

\section{Results and discussion} \label{Sec:4}

    \subsection{Training Parameters and GPU-Usage} \label{Subsec:4.1}
    
\begin{figure}
             \centering
             \begin{subfigure}[b]{0.49\textwidth}
                 \centering
                 \includegraphics[width=\textwidth]{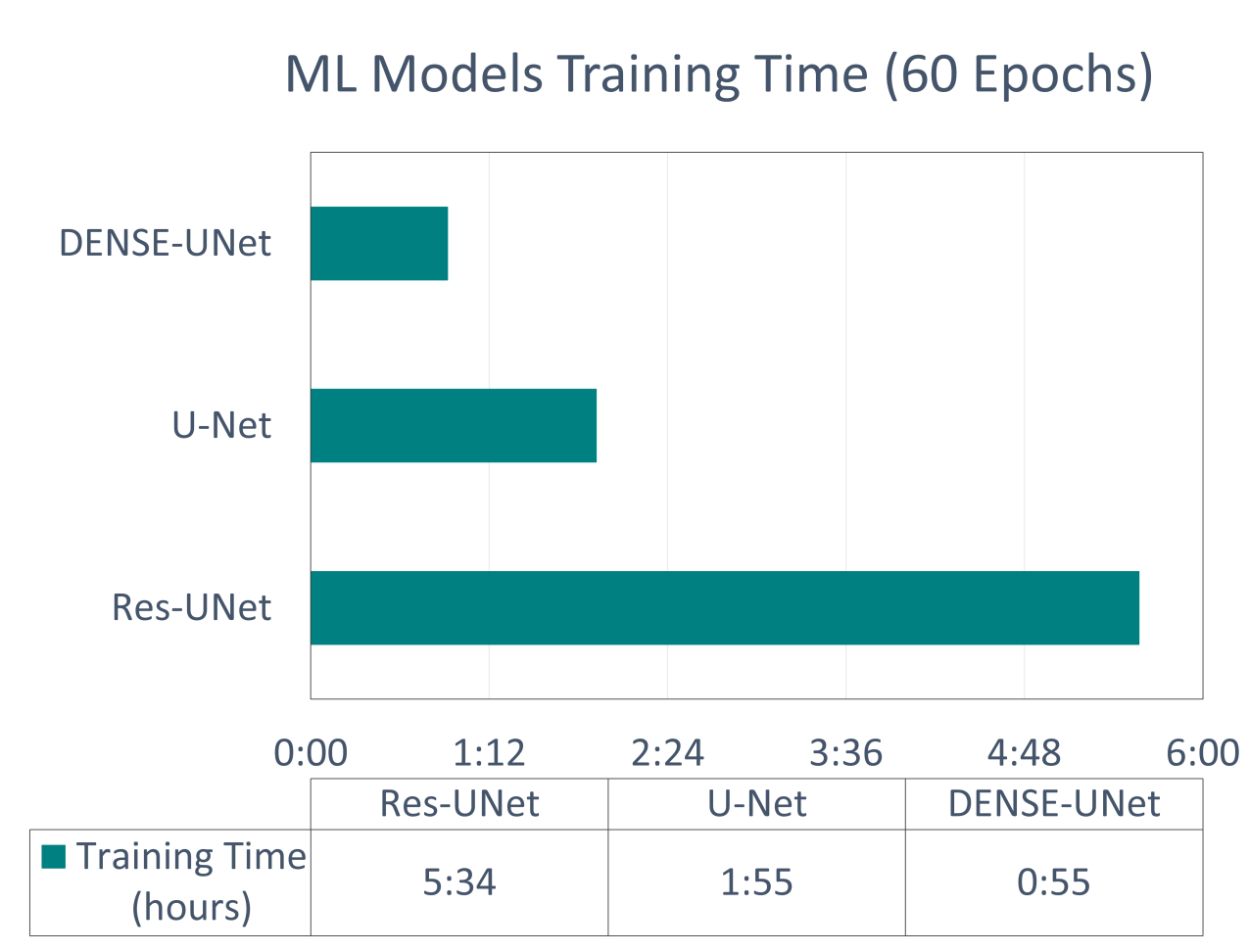}
                 \caption{Total number of trainable parameters}
                 \label{subfig:Parameters}
             \end{subfigure}
             \hfill
             \begin{subfigure}[b]{0.49\textwidth}
                 \centering
                 \includegraphics[width=\textwidth]{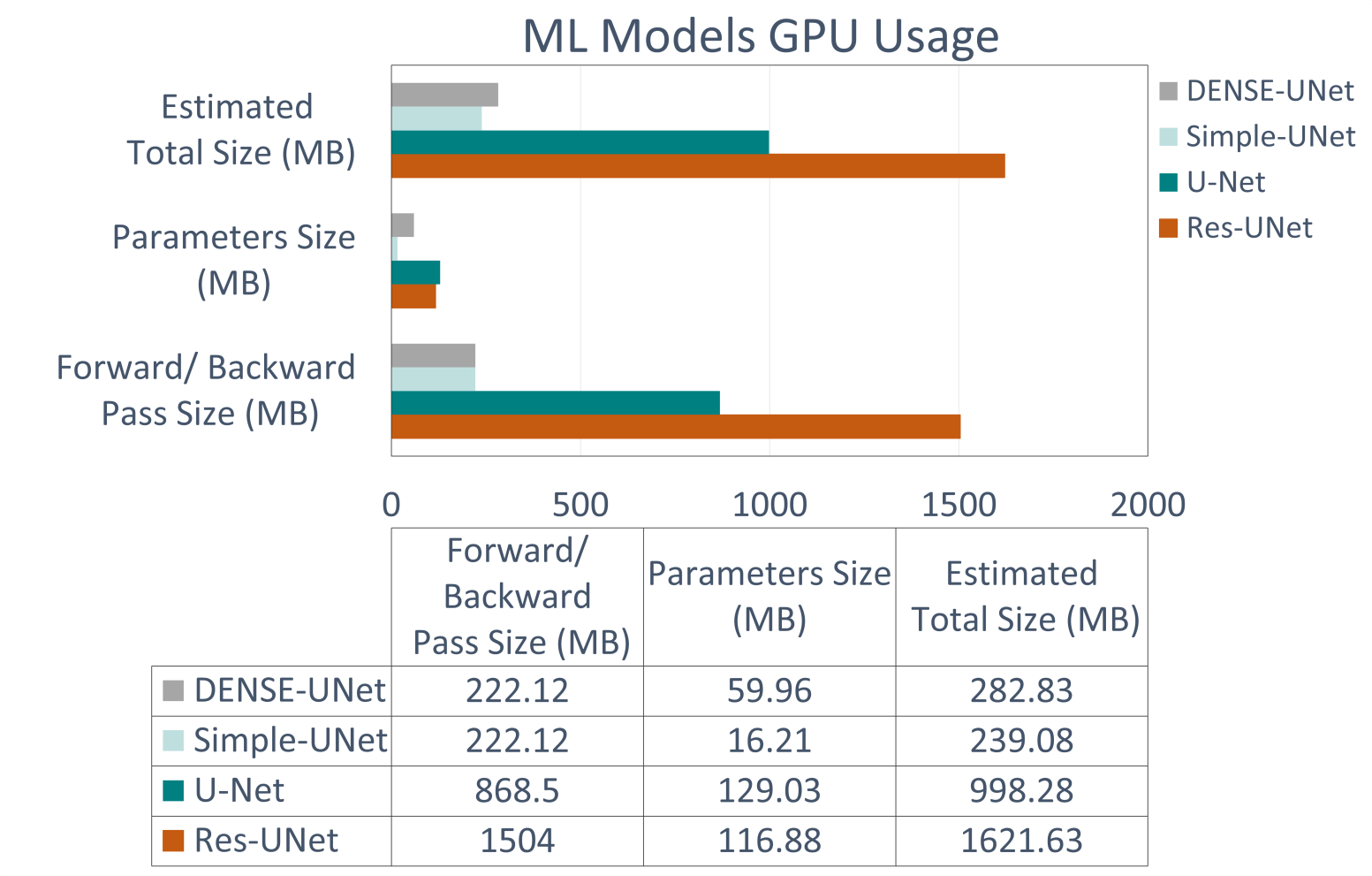}
                 \caption{Total GPU Usage (MB)}
                 \label{subfig:GPU_Usage}
             \end{subfigure}
                \caption{Total number of training parameters, and GPU usage of ML encoder-decoder networks}
                \label{fig:Parameters_and_GPU}
        \end{figure}

        We compared the total number of training parameters and total GPU usage (MBs) of all the ML encoder-decoder networks (U-Net, Res-UNet, DENSE-UNet, and Simple-UNet) used in this framework. 
        From Figure \ref{subfig:Parameters}, we observe that the conventional U-Net model had the highest number of training parameters with 33,824,003 parameters, followed by Res-UNet with 30,639,171 parameters. 
        The model with the least number of training parameters was the Simple-UNet with 4,249,309.
        We note that the the Simple-UNet was designed only for binary segmentation of pores, or particles.
        For RGB segmentation of grains, we used the optimized DENSE-UNet with only 15,718,109 training parameters.

        From Figure \ref{subfig:GPU_Usage}, we note that while the conventional U-Net model contains the largest number of training parameters, the Res-UNet requires a larger memory usage of 1621.63 MB, compared to the U-Net of 998.28 MB.
        This is due to the fact that the forward/backward pass during training is significantly higher for the Res-UNet of 1504 MB, than for the U-Net of 868.5 MB.
        The optimized DENSE-UNet model shows a substantial improvement in GPU memory requirements of 282.83 MB compared to the conventional networks used in RGB segmentation.
        Finally, the Simple-UNet used for binary segmentation of pores, or particles shows the least GPU usage requirements of 239.8 MB.

    \subsection{Framework's output on different microstructures} \label{Subsubsec:4.2}
        \subsubsection{Powder particles and pores} \label{Subsubsec:4.2.1}
        
\begin{figure}
                \centering
                \begin{subfigure}[c]{0.25\textwidth}
                    \centering
                    \begin{subfigure}[t]{1\textwidth}    
                       \centering \includegraphics[width=0.7\linewidth]{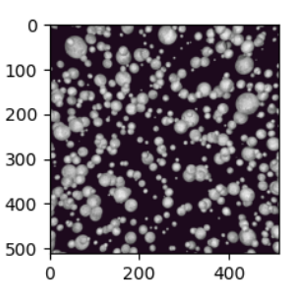}
                        \caption{Input 1}
                    \end{subfigure}
                    
                    \begin{subfigure}[b]{1\textwidth}    
                       \centering
                        \includegraphics[width=0.7\linewidth]{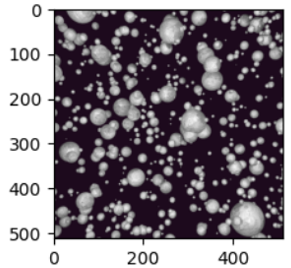}
                        \caption{Input 2}
                    \end{subfigure}
                \end{subfigure}
                \begin{subfigure}[c]{0.25\textwidth}
                    \centering
                    \begin{subfigure}[t]{1\textwidth}    
                      \centering
                       \includegraphics[width=0.7\linewidth]{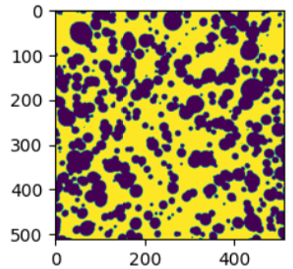}
                        \caption{WCBD: Segmentation}
                    \end{subfigure}
                    
                    \begin{subfigure}[b]{1\textwidth}    
                       \centering
                        \includegraphics[width=0.7\linewidth]{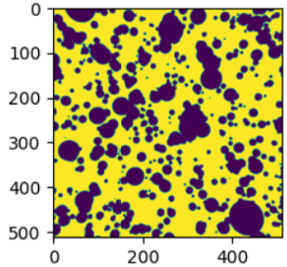}
                        \caption{WCBD: Segmentation}
                    \end{subfigure}
                \end{subfigure}
                \begin{subfigure}[c]{0.25\textwidth}
                    \centering
                    \begin{subfigure}[t]{1\textwidth}    
                       \centering
                        \includegraphics[width=0.7\linewidth]{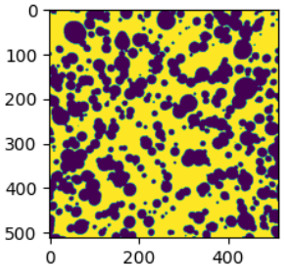}
                        \caption{Simple-UNet: Segmentation}
                    \end{subfigure}
                    
                    \begin{subfigure}[b]{1\textwidth}    
                       \centering
                        \includegraphics[width=0.7\linewidth]{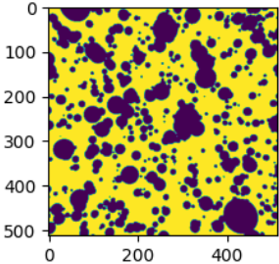}
                        \caption{Simple-UNet: Segmentation}
                    \end{subfigure}
                \end{subfigure}
                    \caption{Comparison of particle's binary segmentation using the watershed and city-block distance segmentation method versus using the ML Simple-UNet}
                    \label{fig:Paricle_Segmentation_Comparison}
            \end{figure}

            We compared the performance of the Simple-UNet on the binary segmentation of synthetic powder particles 
            against the conventional WCBD algorithm as shown in Figure \ref{fig:Paricle_Segmentation_Comparison}.
            We observed that for two distinct cases of varying PSD, the output by the Simple-UNet was qualitatively nearly identical to the segmentations obtained from the conventional method.
        
            For quantitative validation,
            {
            given that the purpose of the Simple-UNet network was to generate accurate binary segmentations which would then be used as the input to the YOLOv5 network,
            }
            we computed the average pixel error and maximum pixel error \cite{Ronneberger2015unet}. 
            We performed a total of 10 segmentation tests for 10 synthetic particle microstructures.
            The resulting average error was 0.15 $\%$ for the predicted ``0'' pixels (black), and 0.06 $\%$ for ``1'' pixels (white).
            Additionally, the maximum computed pixel error was 0.65 $\%$ for ``0'' pixels, and 0.19 $\%$ for ``1'' pixels.
            Given a maximum pixel error less than 1 $\%$, we conclude that the Simple-UNet performs  binary segmentations for synthetic particles with good accuracy.
            
\begin{figure}
                \centering
                \begin{subfigure}[c]{0.30\textwidth}
                    \centering
                    \begin{subfigure}[t]{1\textwidth}    
                       \centering \includegraphics[width=0.80\linewidth]{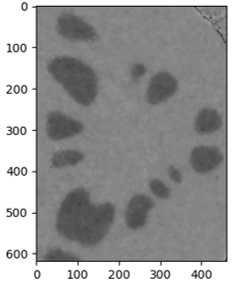}
                        \caption{Input: Pores}
                        \label{subfig:input_pore}
                    \end{subfigure}
                    
                    \begin{subfigure}[b]{1\textwidth}    
                       \centering
                        \includegraphics[width=0.80\linewidth]{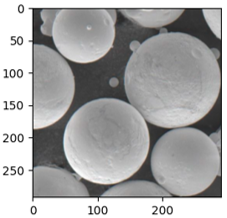}
                        \caption{Input: Particles}
                        \label{subfig:input_particle}
                    \end{subfigure}
                \end{subfigure}
                \begin{subfigure}[c]{0.30\textwidth}
                    \centering
                    \begin{subfigure}[t]{1\textwidth}    
                      \centering
                       \includegraphics[width=0.75\linewidth]{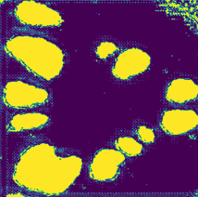}
                        \caption{Simple-UNet: Segmentation}
                        \label{subfig:seg_pore}
                    \end{subfigure}
                    
                    \begin{subfigure}[b]{1\textwidth}    
                       \centering
                        \includegraphics[width=0.75\linewidth]{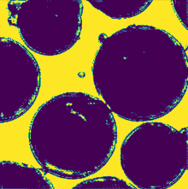}
                        \caption{Simple-UNet: Segmentation}
                        \label{subfig:seg_particle}
                    \end{subfigure}
                \end{subfigure}
                \begin{subfigure}[c]{0.30\textwidth}
                    \centering
                    \begin{subfigure}[t]{1\textwidth}    
                       \centering
                        \includegraphics[width=0.75\linewidth]{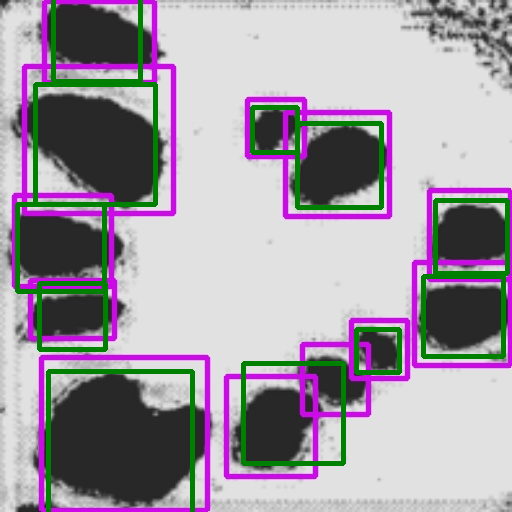}
                        \caption{WCBD (Green) VS YOLOv5 (Purple) bounding boxes}
                        \label{subfig:Yolo_pore}
                    \end{subfigure}
                    
                    \begin{subfigure}[b]{1\textwidth}    
                       \centering
                        \includegraphics[width=0.75\linewidth]{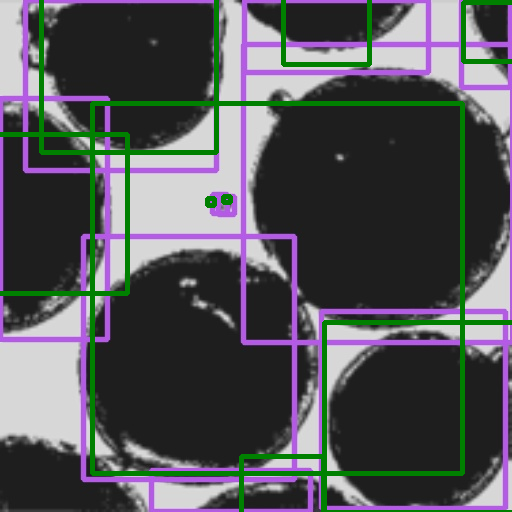}
                        \caption{WCBD (Green) VS YOLOv5 (Purple) bounding boxes}
                        \label{subfig:Yolo_particle}
                    \end{subfigure}
                \end{subfigure}
                \caption{ML framework tested on different distributions of microstructural images of pores and particles.}
                \label{fig:particle_pore_dist}
            \end{figure}

            As an extra step, we validated the framework's performance on material microstructures of pores and particles obtained from different distributions.
            For this, we chose two material samples of austenite, and Ti-6Al-4V, as shown in Figure \ref{fig:particle_pore_dist}.
            We performed binary segmentation using the proposed Simple-UNet, and used YOLOv5 network to obtain bounding boxes for each feature. 
            From Figures \ref{subfig:Yolo_pore} and \ref{subfig:Yolo_particle}, the YOLOv5's predicted bounding boxes are shown in purple, and the conventional WCBD's bounding boxes are shown in green, for pores and particles, respectively.
            Qualitatively, we observed that in both materials the conventional WCBD's boxes are typically smaller in size and of square geometry.
            This is because the WCBD method detects a diameter value for each defect's height and width, while the YOLOv5 network generates rectangular bounding boxes of varying heights and widths for each defect.


            We present the quantitative comparison in Tables \ref{table:2}, 
            \ref{table:3} and \ref{table:4}.
            First, we generated the ground truth bounding boxes of austenite and Ti-6Al-4V using the online open source label tool "CVAT.org".   
            Next, we compared the total number of defects (particles and pores) captured by the WCBD and YOLOv5 with the ground truth
            {
                as shown in Table \ref{table:2}
            }
            . Comparing the ground truth number of pores 
            (total of 12 pores) to Figure \ref{subfig:Yolo_pore}, we confirm that a total number of 12 pores were correctly recognized by the YOLOv5 network, while the WCBD method recognized only 11 pores.
            Comparing the ground truth of 10 particles 
            to Figure \ref{subfig:Yolo_particle}, both the YOLOv5 network (9 predicted) and the WCBD method (8 predicted) fail to capture the particle located at the bottom left location. 
            We note that WCBD method recognized the two largest particles as overlapping, resulting in a single bounding box for both defects.
            However, the YOLOv5 network is able to generate separate bounding boxes for each particle, obtaining a closer result to the ground truth. 
            
                 \begin{table}
                    \centering
                    \begin{tabular}{|p{3.5cm}|p{3.8cm}|p{3.8cm}|}
                        \hline
                       \multicolumn{3}{|c|}{Number of Predicted Defects}\\
                         \hline
                         \centering{} &\centering{Number of Particles} &{Number of Pores} \\
                         \hline
                         \centering{Ground Truth} &\centering{10} &{12}\\
                         \hline
                         \centering{WCBD Method} &\centering{8} &{11} \\
                        \hline
                        \centering{YOLOv5 Method} &\centering{9} &{12} \\
                         \hline
                    \end{tabular}
                    \caption{Comparison of the number of detected bounding boxes for powder particles in Ti-6Al-4V, and pores in Austenite, using the YOLOv5 network and conventional WCBD method versus the ground truth.}
                    \label{table:2}
                 \end{table}

            Tables \ref{table:3} and \ref{table:4} show the obtained 
            {
                percentage
            } 
            errors for the WCBD method and the YOLOv5 network, on the resulting bounding boxes' parameters when compared to the ground truth. 
            We computed the errors of the center coordinates, heights, and widths of each defect by taking the distance difference from the ground truth parameters.
            For powder particles in Ti-6Al-4V shown in Table \ref{table:3}, the WCBD method showed higher error for each parameter. 
            This is because the WCBD method captured the larger particles as overlapping defects.
            Hence, since the YOLOv5 network is able to capture each particle with separate bounding boxes, the resulting errors are significantly lower compared to the WCBD method.
            We note that the 17\% error of YOLOv5, while lower than WCBD method, is still high.
            This error may be caused by the YOLOv5 not capturing the particle located at the bottom left of the Ti-6Al-5V microstructure, shown in Figure \ref{subfig:Yolo_particle}.

            \begin{table}
                \centering
                \begin{tabular}{|p{3.5cm}|c|c|c|c|} 
                    \hline
                    \multicolumn{5}{|c|}{Particles Microstructure -  Ti-6Al-4V}\\
                     \hline
                     \multicolumn{5}{|c|}{Bounding Boxes Error} \\
                     \hline
                     \centering{} &\centering{Center (X)} &\centering{Center (Y)} &\centering{Width} &{Height} \\
                     \hline
                     \centering{WCBD $\%$ Error (Avg. $\pm$ Std. Dev.)} &\centering{$5.45 \pm 7.88$} &\centering{$13.29 \pm 28.43$} &\centering{$44.76\pm 28.22$} &{$51.19\pm 34.20$} \\
                     \hline
                     \centering{YOLO $\%$ Error (Avg. $\pm$ Std. Dev.)} &\centering{$3.34 \pm 4.97$} &\centering{$10.15 \pm 28.80$} &\centering{$17.42 \pm 30.46$} &{$19.59 \pm 29.94$} \\
                     \hline
                \end{tabular}
                \caption{Percentage error for the predicted bounding boxes parameters (centroid locations, height, and width) in powder particles of Ti-6Al-4V, using the YOLOv5 network and the conventional WCBD method compared to the ground truth.}
                \label{table:3}
            \end{table}
            
            For porous defects in austenite, shown in Table \ref{table:4}, the percent error of the WCBD method also shows to be the highest for the centroid coordinates, and width parameters. 
            However, the height $\%$ error of the YOLOv5 was higher than the conventional WCBD method. 
            The center coordinates predicted by the YOLOv5 network showed the highest accuracy with approximately 3.2 $\%$ pixel error, compared to the WCBD method of approximately 10.5 $\%$ pixel error.
            We conclude that the YOLOv5 network was not only able to predict a larger number of defects for porous and powder particle microstructures, but was able to predict the center coordinates of the bounding boxes with higher precision.

            \begin{table}
                \centering
                \begin{tabular}{|p{3.5cm}|c|c|c|c|}
                    \hline
                    \multicolumn{5}{|c|}{Pores Microstructure -  Austenite}\\
                     \hline
                     \multicolumn{5}{|c|}{Bounding Boxes Error} \\
                     \hline
                     \centering{} &\centering{Center (X)} &\centering{Center (Y)} &\centering{Width} &{Height} \\
                     \hline
                     \centering{WCBD $\%$ Error (Avg. $\pm$ Std. Dev.)} &\centering{$6.06 \pm 18.58$} &\centering{$6.79 \pm 21.31$} &\centering{$27.46\pm 25.05$} &{$18.82\pm 27.30$} \\
                     \hline
                     \centering{YOLO $\%$ Error (Avg. $\pm$ Std. Dev.)} &\centering{$0.59 \pm 0.59$} &\centering{$0.78 \pm 0.58$} &\centering{$11.74 \pm 7.39$} &{$22.52 \pm 12.15$} \\
                     \hline
                \end{tabular}
                \caption{Percentage error for the predicted bounding boxes parameters (centroid locations, height, and width) in pores of Austenite, using the YOLOv5 network and the conventional WCBD method compared to the ground truth.}
                \label{table:4}
            \end{table}

        \subsubsection{Grain Boundaries} \label{Subsubsec:4.2.2}    
        
\begin{figure}
                \centering
                \begin{subfigure}[c]{0.19\textwidth}
                    \centering
                    \begin{subfigure}[c]{1\textwidth}    
                       \centering \includegraphics[width=0.7\linewidth]{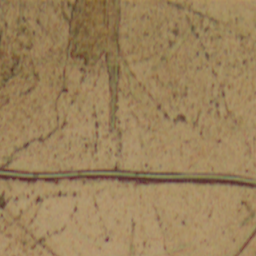}
                        \caption{Input Image}
                        \label{subfig:input_grain_same}
                    \end{subfigure}
                \end{subfigure}
                \begin{subfigure}[c]{0.19\textwidth}
                    \centering
                    \begin{subfigure}[c]{1\textwidth}    
                        \centering
                        \includegraphics[width=0.7\linewidth]{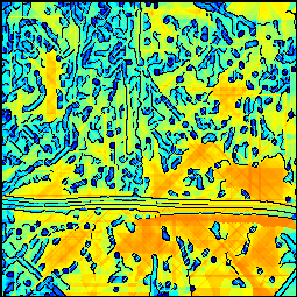}
                        \caption{PSILM}
                        \label{subfig:Matlab_grain_same}
                    \end{subfigure}
                \end{subfigure}
                \begin{subfigure}[c]{0.19\textwidth}
                    \centering
                    \begin{subfigure}[c]{1\textwidth}    
                      \centering                       
                      \includegraphics[width=0.7\linewidth]{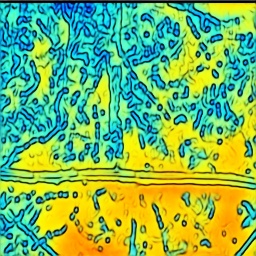}
                        \caption{Res-UNet}
                        \label{subfig:Res-UNet_grain_same}
                    \end{subfigure}
                \end{subfigure}
                \begin{subfigure}[c]{0.19\textwidth}
                    \centering
                    \begin{subfigure}[c]{1\textwidth}    
                      \centering
                       \includegraphics[width=0.7\linewidth]{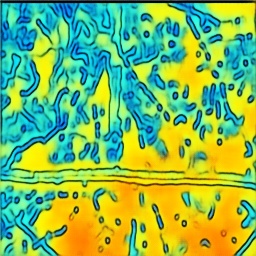}
                        \caption{U-Net}
                        \label{subfig:UNet_grain_same}
                    \end{subfigure}
                \end{subfigure}
                \begin{subfigure}[c]{0.19\textwidth}
                    \centering
                    \begin{subfigure}[c]{1\textwidth}    
                      \centering
                       \includegraphics[width=0.7\linewidth]{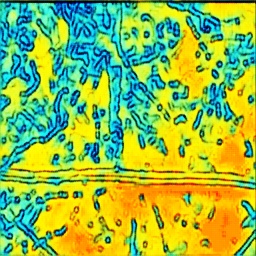}
                        \caption{DENSE-UNet
                        }
                        \label{subfig:DENSE_grain_same}
                    \end{subfigure}
                \end{subfigure}
                    \caption{Comparison of grain boundaries' RGB segmentation output from ASTM's PSILM, Res-UNet, U-Net, and DENSE-UNet.}
                    \label{fig:seg_same}
            \end{figure}

            We compared the performance of the ML encoder-decoder networks used for RGB segmentation of GBs (Res-UNet, U-Net, and DENSE-UNet) against the ASTM's PSILM as shown in Figure \ref{subfig:input_grain_same}.
            We present the resultant GB segmentations obtained for the Res-UNet, U-Net, and DENSE-UNet in Figures \ref{subfig:Res-UNet_grain_same}, \ref{subfig:UNet_grain_same}, and \ref{subfig:DENSE_grain_same}, respectively.
            
            Qualitatively, we note that the ML encoder-decoder networks perform accurate segmentations for the grains' edges, capturing the larger grain sections as color variations from yellow to dark red, to the smaller grain sections as green to dark blue.
            In contrast, the PSILM algorithm does not fully capture some of the larger sized grains shown in Figure \ref{subfig:input_grain_same}.
            This may be a cause of the PSILM requiring the use of conventional edge detection image-processing tools.
            For example, following the edge detection, a random number points are placed overlapping the detected edges in order to calculate the lengths between each point using the line intercept method \cite{Lehto2016Characterization, Lehto2014Influence}.
            These calculated distances are then interpolated
            to generate the color variations.
            However, conventional edge detection tools may occasionally result in some of the lighter shadows to be incorrectly labeled as small grains \cite{Kalra2016Canny}, omitting some of the larger grains.
            Additionally, the line intercept calculations do not take into account the edges of the domain.
            As such, 
            some of the larger grains located at the corners of the microstructure may be incorrectly segmented as smaller grains.

            \begin{figure}
                \centering
                \begin{subfigure}[c]{0.25\textwidth}
                    \centering
                    \begin{subfigure}[c]{1\textwidth}    
                       \centering \includegraphics[width=0.7\linewidth]{4_train.png}
                        \caption{Grain Boundary}
                    \end{subfigure}
                \end{subfigure}
                \begin{subfigure}[c]{0.25\textwidth}
                    \centering
                    \begin{subfigure}[t]{1\textwidth}    
                      \centering
                       \includegraphics[width=0.7\linewidth]{MATLAB_Same.png}
                        \caption{PSILM Output}
                        \label{subfig:MATLAB_seg_same}
                    \end{subfigure}
                    
                    \begin{subfigure}[b]{1\textwidth}    
                       \centering
                       \includegraphics[width=0.7\linewidth]{DENSE_UNet_Same.jpg}
                        \caption{DENSE-UNet Output}
                        \label{subfig:DENSE_seg_same}
                    \end{subfigure}
                \end{subfigure}
                \begin{subfigure}[c]{0.48\textwidth}
                    \centering
                    \begin{subfigure}[c]{1\textwidth}    
                       \centering
                        \includegraphics[width=1\linewidth]{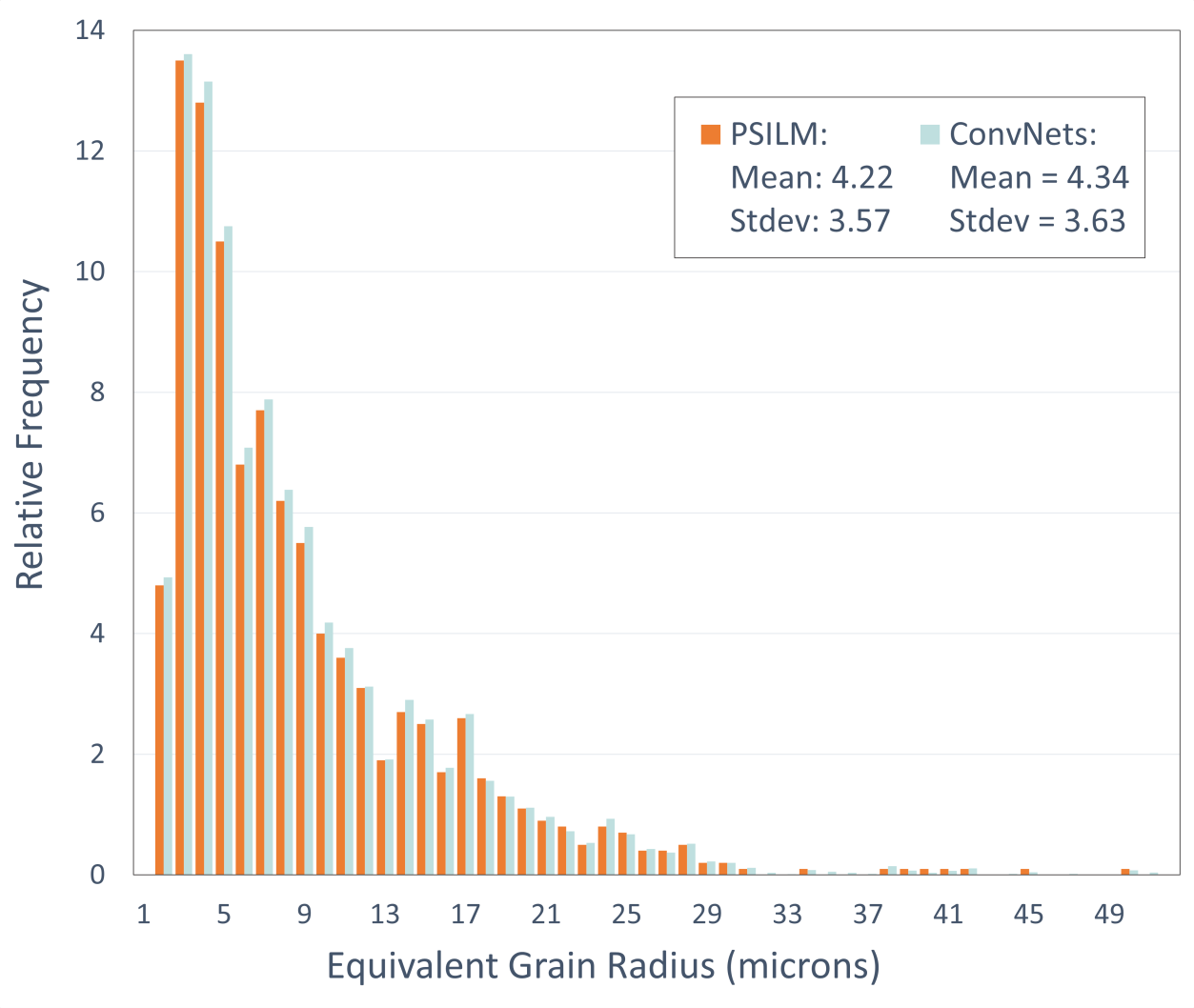}
                        \caption{PSILM vs ML Framework: Grain boundary size distribution histograms}
                    \end{subfigure}
                \end{subfigure}
                    \caption{Comparison of grain boundaries' size distribution histograms from the PSILM, and the ML framework.}
                    \label{fig:seg_hist}
            \end{figure}
            
            In Figure \ref{fig:seg_hist}, the PSILM's histograms were compared to the framework's predicted histograms.
            {
                The mean and standard deviation of the grain radius as predicted by the PSILM and CNNs algorithms are $4.22 \pm 3.57$, and $4.34 \pm 3.63$ respectively.
            }
            We observe that while the histogram from the ML framework is nearly identical to ASTM's PSILM output, the radius distribution predicted by the ML framework is slightly higher.
            This may be caused by the aforementioned consequences from the PSILM method when using image-processing edge detection tools and interpolation functions.
            However, the DENSE-UNet is able to capture a higher amount of the larger grain sections, causing the histogram's output to be slightly higher when compared to the PSILM.
            
\begin{figure}
                \centering
                \begin{subfigure}[c]{0.27\textwidth}
                    \centering
                    \begin{subfigure}[t]{1\textwidth}    
                       \centering \includegraphics[width=0.7\linewidth]{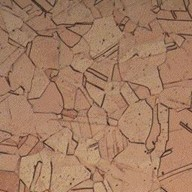}
                        \caption{Input: Copper}
                        \label{subfig:Copper_Alloy}
                    \end{subfigure}
                    
                    \begin{subfigure}[c]{1\textwidth}    
                       \centering
                        \includegraphics[width=0.7\linewidth]{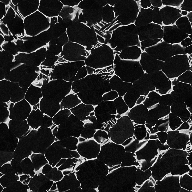}
                        \caption{Input: Titanium}
                        \label{subfig:Titanium}
                    \end{subfigure}
                    
                    \begin{subfigure}[c]{1\textwidth}    
                       \centering
                        \includegraphics[width=0.7\linewidth]{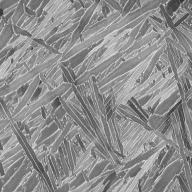}
                        \caption{Input: Inconel Alloy}
                        \label{subfig:Inconel}
                    \end{subfigure}
                    
                    \begin{subfigure}[b]{1\textwidth}    
                       \centering
                        \includegraphics[width=0.7\linewidth]{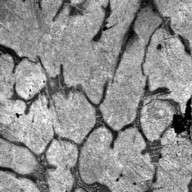}
                        \caption{Input: Steel}
                        \label{subfig:Steel}
                    \end{subfigure}
                \end{subfigure}
                \begin{subfigure}[c]{0.27\textwidth}
                    \centering
                    \begin{subfigure}[t]{1\textwidth}    
                       \centering \includegraphics[width=0.7\linewidth]{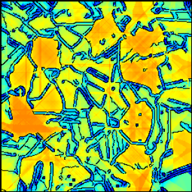}
                        \caption{PSILM: Copper}
                        \label{subfig:MATLAB_Copper}
                    \end{subfigure}
                    
                    \begin{subfigure}[c]{1\textwidth}    
                       \centering
                        \includegraphics[width=0.7\linewidth]{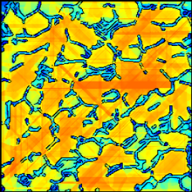}
                        \caption{PSILM: Titanium}
                        \label{subfig:MATLAB_Titanium}
                    \end{subfigure}
                    
                    \begin{subfigure}[c]{1\textwidth}    
                       \centering
                        \includegraphics[width=0.7\linewidth]{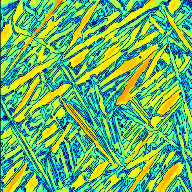}
                        \caption{PSILM: Inconel}
                        \label{subfig:MATLAB_Inconel}
                    \end{subfigure}
                    
                    \begin{subfigure}[b]{1\textwidth}    
                       \centering
                        \includegraphics[width=0.7\linewidth]{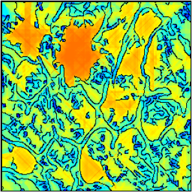}
                        \caption{PSILM: Steel}
                        \label{subfig:MATLAB_Steel}
                    \end{subfigure}
                \end{subfigure}
                \begin{subfigure}[c]{0.27\textwidth}
                    \centering
                    \begin{subfigure}[t]{1\textwidth}    
                       \centering
                      
                      \includegraphics[width=0.7\linewidth]{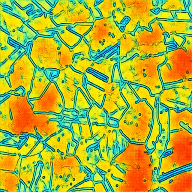} 
                        \caption{DENSE-UNet: Copper}
                        \label{subfig:DENSE_Copper}
                    \end{subfigure}
                    
                    \begin{subfigure}[c]{1\textwidth}    
                       \centering
                        \includegraphics[width=0.7\linewidth]{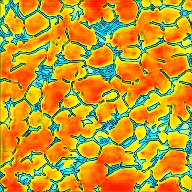}
                        \caption{DENSE-UNet: Titanium}
                        \label{subfig:DENSE_Titanium}
                    \end{subfigure}
                    
                    \begin{subfigure}[c]{1\textwidth}    
                       \centering
                        \includegraphics[width=0.7\linewidth]{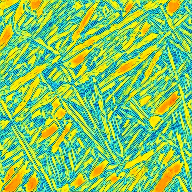}
                        \caption{DENSE-UNet: Inconel}
                        \label{subfig:DENSE_Inconel}
                    \end{subfigure}
                    
                    \begin{subfigure}[b]{1\textwidth}    
                       \centering
                        \includegraphics[width=0.7\linewidth]{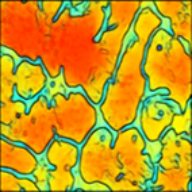}
                        \caption{DENSE-UNet: Steel}
                        \label{subfig:DENSE_Steel}
                    \end{subfigure}
                \end{subfigure}
                    \caption{ML framework tested on different distributions of microstructural images of grain boundaries.}
                    \label{fig:seg_dist}
            \end{figure}

            Next, we tested the performance of DENSE-UNet's RGB segmentation 
            on multiple materials containing microstructural GBs.
            We obtained 
            four microstructural images of GBs for a Copper Alloy, Titanium, an Inconel Alloy, and Steel as shown in Figures \ref{subfig:Copper_Alloy}, \ref{subfig:Titanium}, \ref{subfig:Inconel}, and \ref{subfig:Steel}, respectively. 
            We obtained their RGB segmentations using the ASTM's PSILM algorithm as shown in Figures \ref{subfig:MATLAB_Copper}, \ref{subfig:MATLAB_Titanium}, \ref{subfig:MATLAB_Inconel}, and \ref{subfig:MATLAB_Steel}.
            Next, we generated RGB segmentations using the DENSE-UNet as shown in Figures \ref{subfig:DENSE_Copper}, \ref{subfig:DENSE_Titanium}, \ref{subfig:DENSE_Inconel}, and \ref{subfig:DENSE_Steel}.
            We observed that the PSILM algorithm incorrectly labeled some of the larger sized grains as small in certain sections of the microstructure (specifically the corner areas).
            This may be seen by comparing Figures \ref{subfig:MATLAB_Copper} to \ref{subfig:DENSE_Copper}, and \ref{subfig:MATLAB_Titanium} to \ref{subfig:DENSE_Titanium}, for Copper and Titanium, respectively. 

            The difference is more evident for steel as shown in Figure \ref{subfig:MATLAB_Steel} and \ref{subfig:DENSE_Steel}.
            For instance, some areas involving larger grains are incorrectly labeled as smaller grains by PSILM algorithm.
            Additionally, the algorithm predicts smaller grains within larger sized grains.
            In contrast, the DENSE-UNet's segmentation for steel shown in Figure \ref{subfig:DENSE_Steel}, captures the larger grains with higher accuracy than the conventional method. 
            By comparing Figures \ref{subfig:MATLAB_Steel} and \ref{subfig:DENSE_Steel}, we concluded that the optimized encoder-decoder network implemented in the ML framework (DENSE-UNet), is able to accurately generalize the relationship that exists when segmenting larger grains as yellow-dark red color variations, to smaller grains as green-dark blue color variations.
            
    \subsection{ML encoder-decoders training time and accuracy} \label{Subsec:4.3}
    
\begin{figure}
             \centering
             \begin{subfigure}[b]{0.49\textwidth}
                 \centering
                 \includegraphics[width=\textwidth]{Train_Time_TIF}
                 \caption{Training time (min:sec) tested for 60 Epochs}
                 \label{fig:TrainTime}
             \end{subfigure}
             \hfill
             \begin{subfigure}[b]{0.49\textwidth}
                 \centering
                 \includegraphics[width=\textwidth]{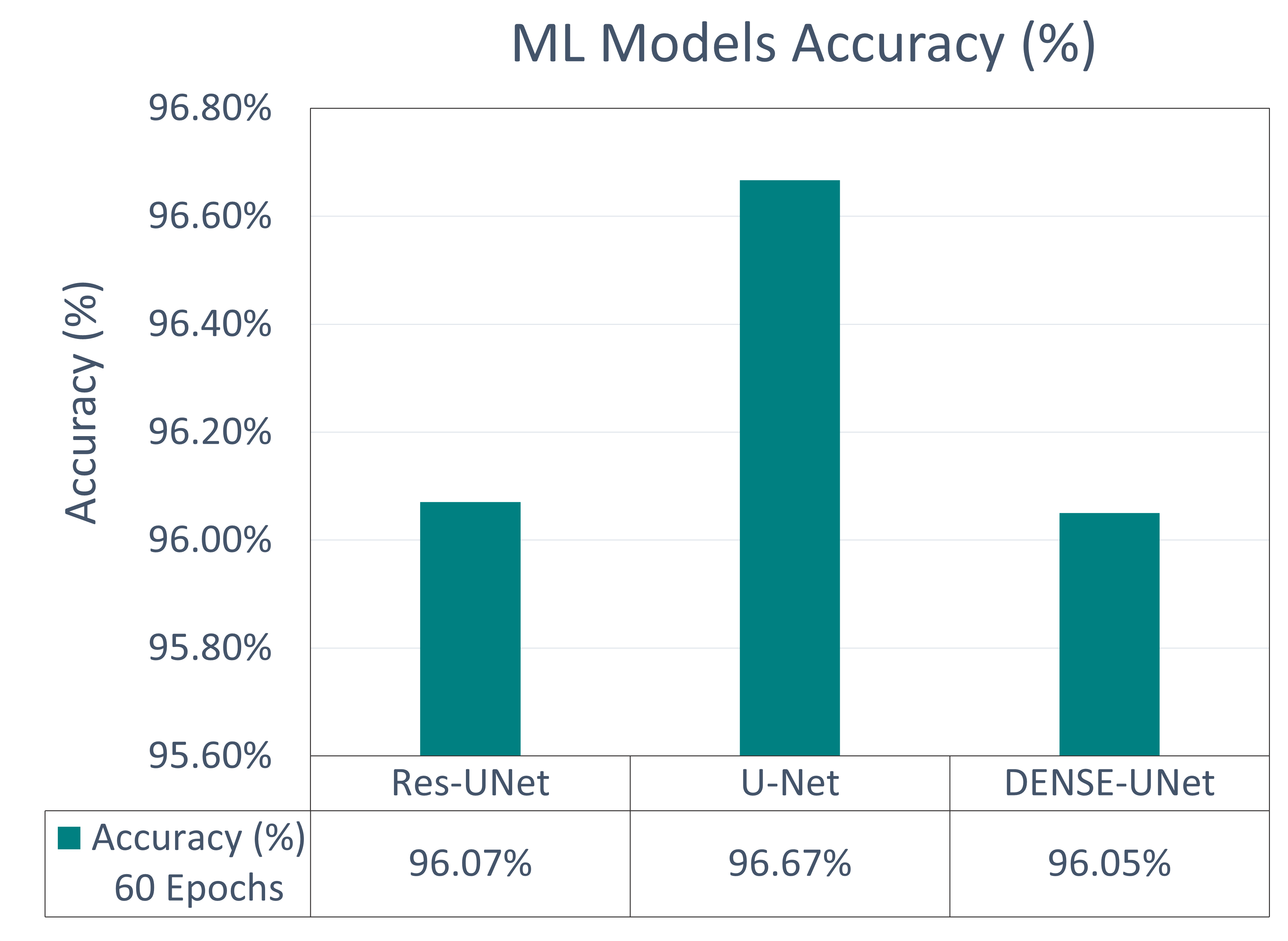}
                 \caption{Accuracy ($\%$) tested for 60 epochs}
                 \label{fig:Accuracy}
             \end{subfigure}
                \caption{Training time, and accuracy comparisons of ML encoder-decoder networks}
                \label{Train_and_Accuracy}
        \end{figure}

        We tested the required training times of the encoder-decoder networks used for RGB segmentation of grains (Figure \ref{fig:TrainTime}).
        As expected, the model with the longest training time of 5:34 hours was the Res-UNet due to the highest GPU usage seen in Figure \ref{subfig:GPU_Usage}.
        Similarly, the following model with the longest training time was the U-Net at 3:55 hours.
        We observe that due to its low GPU usage, the optimized DENSE-UNet shows a substantial training time improvement of 55 minutes compared to the conventional RGB segmentation models. 

        Next, we tested the accuracy of the ML encoder-decoder networks for the segmentation of grains (Figure \ref{fig:Accuracy}).
        Interestingly, we observe that although the Res-UNet involves a higher GPU-Usage than the U-Net, its segmentation accuracy is less than the U-Net by 0.60$\%$. 
        This may be a result of the Res-UNet using pre-trained weights, compared to the U-Net which involves a higher number of training parameters as shown in Figure \ref{subfig:Parameters}.
        Therefore, the network with the highest accuracy was the conventional U-Net at 96.67$\%$.
        We emphasize that while the U-Net shows the highest accuracy, the DENSE-UNet's accuracy is nearly identical to the Res-UNet and U-Net at an accuracy of 96.05$\%$.
        By contrast, DENSE-UNet network demonstrates the lowest training time, 4:39 hours less than the Res-UNet, and 1 hour less than the U-Net, making it the most efficient segmentation network of the three (Figure \ref{fig:TrainTime}).

    \subsection{Time performance of the integrated framework versus PSILM} \label{Subsec:4.4}
        
\begin{figure}
            \centering
            \includegraphics[width=0.6\linewidth]{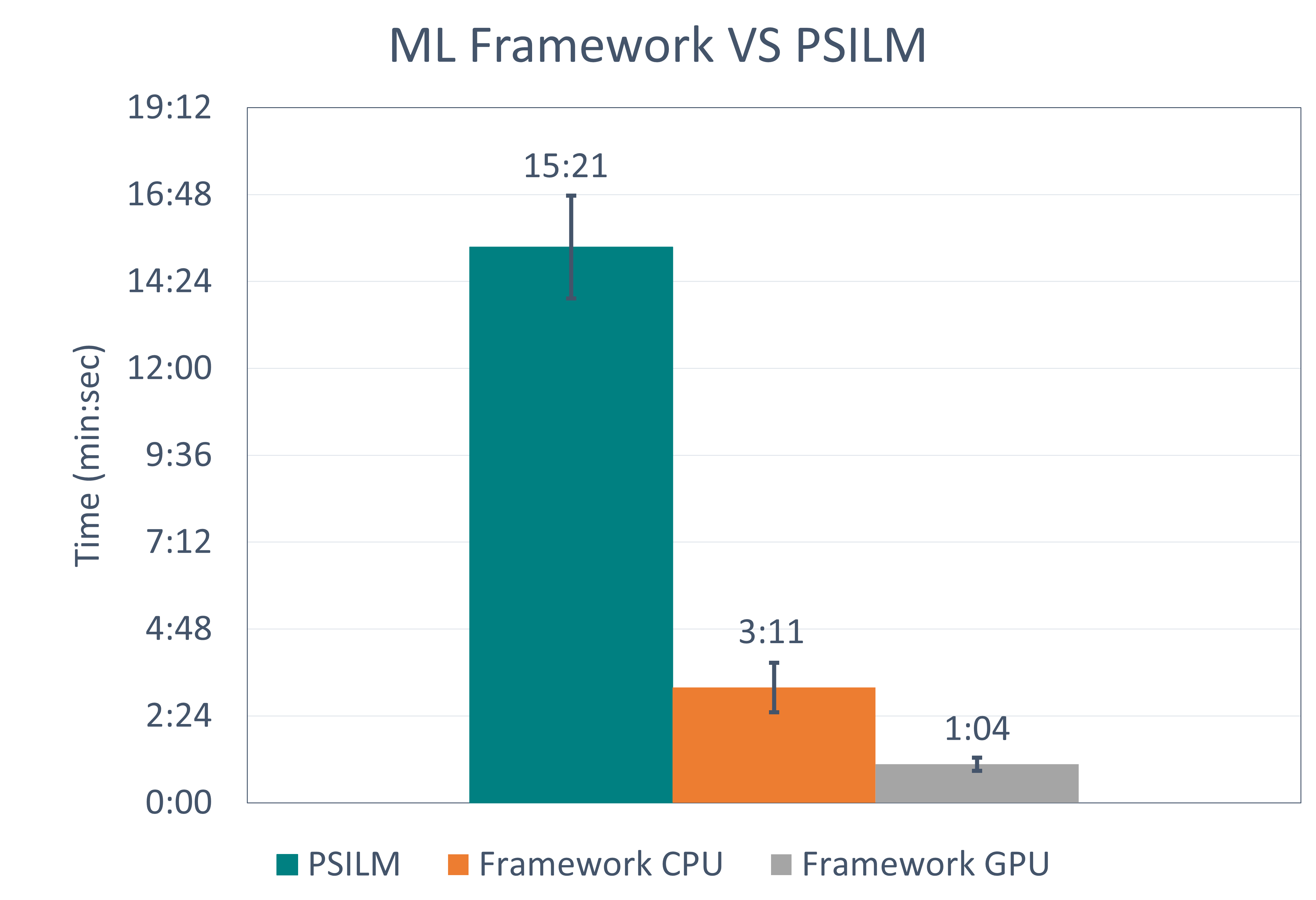}
            \centering
            \caption{Comparison of analysis time of the ML Framework versus the PSILM, for 10 microstructural grain boundary images}
            \label{fig:Framework_vs_MATLAB}
        \end{figure}

        Finally, we compared the time required by the developed framework to characterize 10 microstructural GB images against that of PSILM algorithm (Figure \ref{fig:Framework_vs_MATLAB}).
        We obtained the 
        average characterization time and the standard deviation of each method.
        For the developed framework, we recorded the time from the initial loading of the pre-trained weights for the Classifier CNN, the Simple-UNet, the DENSE-UNet, and the Regression CNNs, to the final outputted histograms containing the grain size distributions.
        We tested the framework's performance using GPU, as well as CPU.
        Similarly, for the PSILM analysis, we recorded the time starting from the input of the image, to the final segmentation and histogram generation. 
        For 10 sample images, the PSILM took 15 minutes and 21 seconds (approximately 1 minute and 32 seconds for each image).
        We observe a substantial time improvement from the framework requiring 3:11 minutes (approximately 19 seconds per image) when using CPU. 
        Additionally, when GPU was utilized, the framework's total analysis time was 1:04 minutes (6.4 seconds per image); significantly faster than the framework's performance using CPU, and the PSILM.

        We emphasize that the PSILM may be advantageous for small datasets when compared to the required training time of the DENSE-UNet. 
        Although ML offers the advantage of saving trained models for future use without the need to re-train each network, when users are required to re-train the models, then the PSILM may provide with faster results.
        For example, to analyze 36 images or fewer, PSILM algorithm requires approximately 55 minutes or less.
        In contrast, to train the DENSE-UNet and generate the GB segmentations and histograms using the framework, would require more than 55 minutes.
        Therefore when training of the networks is necessary, the framework will begin to outperform the conventional method once the number of microstructural images is higher than a certain threshold (36 in this analysis). 
        For instance, for the analysis of 100 microstructural images, the PSILM would require approximately 2 hours and 33 minutes, while the framework with CPU would require 44 minutes to re-train the models, in addition to approximately 32 minutes for the analysis process, resulting in a time improvement of 1 hour and 16 minutes minutes by the framework.
        
        When training is not necessary (pre-trained network), the proposed framework outperforms the conventional segmentation algorithm.
        By noting the PSILM's analysis time of 2:33 hours for 100 images, to the pre-trained framework's analysis time of 32 minutes, a time improvement of approximately 2 hours may be achieved by the framework.
        Lastly, when a mixed dataset of grain boundaries, pores, or particles is required in the analysis, the framework's capability to autonomously classify the resultant microstructure for each defect, could save substantial analysis time.
        Therefore, while the developed framework is not a substitute for existing conventional microstructural feature characterization techniques, specifically, when small datasets are used, it shows to be a promising tool for analyzing large datasets autonomously.
        
    \subsection{Case of microstructures involving multiple defects} \label{Subsec:4.5} 
    {    
            We note that the proposed ML framework may be used to autonomously characterize microstructures involving a single type of defect.
            However, a significant challenge arises for materials involving multiple features within their microstructure.
            To address this issue, we explored the framework's capability to perform this additional task, and propose a future work which may improve it's impact in the complete characterization of various microstructures.
            First, we gathered a new dataset of microstructures including both GBs and pores was gathered. 
            We used the DoITPoMS - Micrograph Library to obtain 354 microstructures by first resizing the raw images to 512x512 for the lower-resolution, 1024x1024 for the higher-resolution, and finally cropping each image to 256x256.
            We also applied data augmentation by randomly rotating, cropping, and flipping vertically or horizontally, to generate a total of 1416 images.
            We implemented the the procedure previously explained in \ref{Subsec:3.1} to obtain the binary segmentation labels, and bounding boxes for each pore in the microstructures.
            We also applied the PSILM method to obtain RGB segmentations for the GBs.
            We note that the PSILM is capable of characterizing the locations of pores using black pixels, excluding them from the grain size distribution measurements.
            Lastly, we applied the transfer learning to the Simple-UNet, DENSE-UNet, and YOLOv5 network on the new microstructures.
            
            As shown in Figure \ref{fig:multiple_binary}, the Simple-UNet qualitatively shows accurate binary segmentations similar to the conventional WCBD method.
            Next, we tested the performance of the YOLOv5 algorithm for the characterization of microstructures involving multiple defects.
            Figure \ref{fig:multiple_yolo} shows the output bounding boxes obtained from the YOLOv5 by inputting the binary segmentations from the Simple-UNet.
            The results shown in Figures \ref{fig:multiple_binary} and \ref{fig:multiple_yolo}, reinforce the framework's ability to fully characterize pores in microstructures of multiple defects.
            The remaining defects to be characterized were GBs.
            Figure \ref{fig:multiple_rgb} shows the ML framework was able to recognize the locations of pores, and the locations of the grain boundaries using the color variation gradients, with similar outputs to the conventional PSILM.
            
            Although these results show future promise to extend the framework's capability for the characterization of multiple types of defects simultaneously, there are additional challenges.
            For example, in order to develop autonomy for these types of cases a larger dataset may be required for training a new Classifier CNN.
            Since the Classifier CNN described in Sections \ref{Subsec:3.1} and \ref{Subsec:3.2} required 2000 microstructural images of pores and particles, and grain boundaries (a total of 4000 images), the new training set of 1239 microstructures falls short of the required amount for an even distribution.
            Therefore, as a future work for the extension of the ML framework, a larger dataset could be generated to train a new Classifier CNN which may detect if the microstructure involves pores, particles, grain boundaries, or both types, in order to then apply the corresponding ML characterization networks. 
                
            \begin{figure}
                \begin{subfigure}[c]{0.32\textwidth}
                    \centering
                    \begin{subfigure}[t]{1\textwidth}    
                       \centering \includegraphics[width=0.7\linewidth]{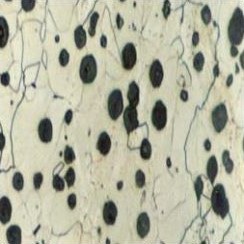}
                        \caption{Input 1}
                    \end{subfigure}
                    
                    \begin{subfigure}[b]{1\textwidth}    
                       \centering
                        \includegraphics[width=0.7\linewidth]{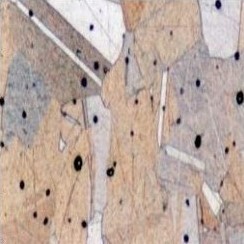}
                        \caption{Input 2}
                    \end{subfigure}
                \end{subfigure}
                \begin{subfigure}[c]{0.32\textwidth}
                    \centering
                    \begin{subfigure}[t]{1\textwidth}    
                      \centering
                       \includegraphics[width=0.7\linewidth]{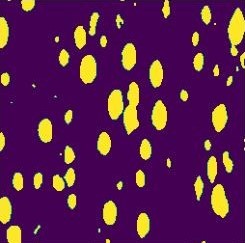}
                        \caption{WCBD: Segmentation}
                    \end{subfigure}
                    
                    \begin{subfigure}[b]{1\textwidth}    
                       \centering
                        \includegraphics[width=0.7\linewidth]{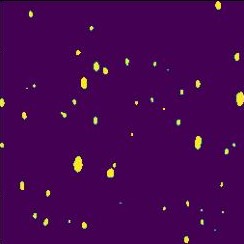}
                        \caption{WCBD: Segmentation}
                    \end{subfigure}
                \end{subfigure}
                \begin{subfigure}[c]{0.32\textwidth}
                    \centering
                    \begin{subfigure}[t]{1\textwidth}    
                       \centering
                        \includegraphics[width=0.7\linewidth]{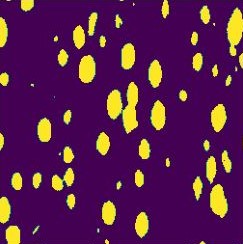}
                        \caption{Simple-UNet: Segmentation}
                    \end{subfigure}
                    
                    \begin{subfigure}[b]{1\textwidth}    
                       \centering
                        \includegraphics[width=0.7\linewidth]{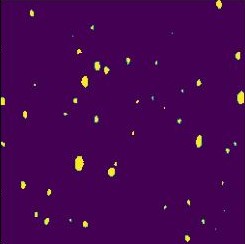}
                        \caption{Simple-UNet: Segmentation}
                    \end{subfigure}
                \end{subfigure}
                \caption{Binary segmentation using WCBD method   versus Simple-UNet for microstructures of grain boundaries and pores}
                    \label{fig:multiple_binary}
            \end{figure}

            \begin{figure}
                \begin{subfigure}[c]{0.32\textwidth}
                    \centering
                    \begin{subfigure}[t]{1\textwidth}    
                       \centering \includegraphics[width=0.7\linewidth]{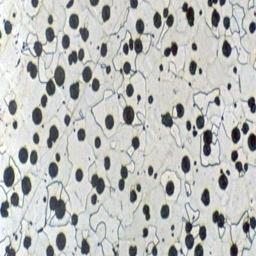}
                        \caption{Input 1}
                    \end{subfigure}
                    
                    \begin{subfigure}[b]{1\textwidth}    
                       \centering
                        \includegraphics[width=0.7\linewidth]{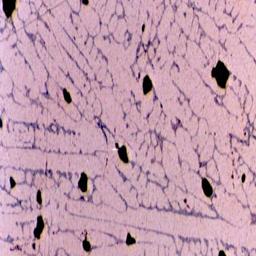}
                        \caption{Input 2}
                    \end{subfigure}
                \end{subfigure}
                \begin{subfigure}[c]{0.32\textwidth}
                    \centering
                    \begin{subfigure}[t]{1\textwidth}    
                      \centering
                       \includegraphics[width=0.7\linewidth]{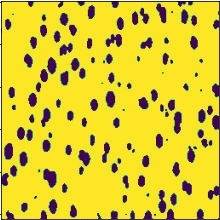}
                        \caption{Simple-UNet: Segmentation}
                    \end{subfigure}
                    
                    \begin{subfigure}[b]{1\textwidth}    
                       \centering
                        \includegraphics[width=0.7\linewidth]{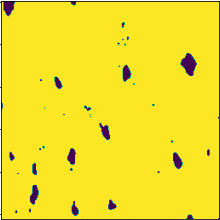}
                        \caption{Simple-UNet: Segmentation}
                    \end{subfigure}
                \end{subfigure}
                \begin{subfigure}[c]{0.32\textwidth}
                    \centering
                    \begin{subfigure}[t]{1\textwidth}    
                       \centering
                        \includegraphics[width=0.7\linewidth]{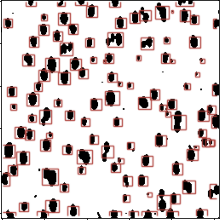}
                        \caption{YOLOv5 Output}
                    \end{subfigure}
                    
                    \begin{subfigure}[b]{1\textwidth}    
                       \centering
                        \includegraphics[width=0.7\linewidth]{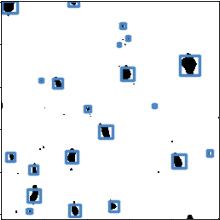}
                        \caption{YOLOv5 Output}
                    \end{subfigure}
                \end{subfigure}
                \caption{YOLOv5's performance on microstructures involving grain boundaries and pores.}
                \label{fig:multiple_yolo}
            \end{figure}

            \begin{figure}
                \begin{subfigure}[c]{0.32\textwidth}
                    \centering
                    \begin{subfigure}[t]{1\textwidth}    
                       \centering \includegraphics[width=0.7\linewidth]{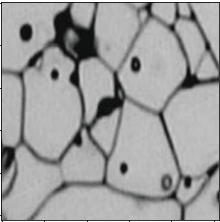}
                        \caption{Input 1}
                    \end{subfigure}
                    
                    \begin{subfigure}[b]{1\textwidth}    
                       \centering
                        \includegraphics[width=0.7\linewidth]{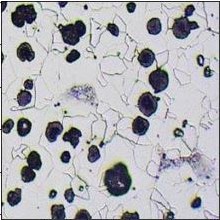}
                        \caption{Input 2}
                    \end{subfigure}
                \end{subfigure}
                \begin{subfigure}[c]{0.32\textwidth}
                    \centering
                    \begin{subfigure}[t]{1\textwidth}    
                      \centering
                       \includegraphics[width=0.7\linewidth]{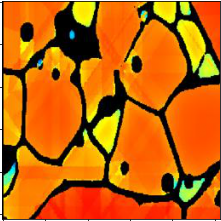}
                        \caption{PSILM: Segmentation}
                    \end{subfigure}
                    
                    \begin{subfigure}[b]{1\textwidth}    
                       \centering
                        \includegraphics[width=0.7\linewidth]{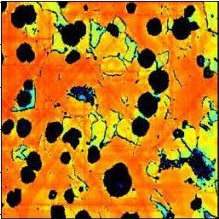}
                        \caption{PSILM: Segmentation}
                    \end{subfigure}
                \end{subfigure}
                \begin{subfigure}[c]{0.32\textwidth}
                    \centering
                    \begin{subfigure}[t]{1\textwidth}    
                       \centering
                        \includegraphics[width=0.7\linewidth]{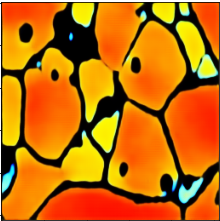}
                        \caption{RGB CEDN: Output}
                    \end{subfigure}
                    
                    \begin{subfigure}[b]{1\textwidth}    
                       \centering
                        \includegraphics[width=0.7\linewidth]{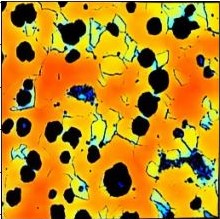}
                        \caption{RGB CEDN: Output}
                    \end{subfigure}
                \end{subfigure}
                \caption{Encoder-Decoder network RGB segmentation performance on microstructures involving grain boundaries and pores.}
                \label{fig:multiple_rgb}
            \end{figure}
     }

\section{Conclusion} \label{Conclusion}

    The proposed ML framework showed promising improvements when compared to conventional methods of microstructural feature extraction, as well as conventional ML encoder-decoder networks used for segmentation.
    By integrating multiple ML algorithms, we achieved a complete and autonomous characterization process for microstructural pores, powder particles, and GBs.
    While conventional computational methods such as the PSILM and the WCBD were necessary to train the ML networks, the finalized framework may autonomously perform material characterization without further training.
    The proposed framework's characterization automation starts by classifying the microstructural defects involved as pores, particles, or grains.
    By using two developed ML encoder-decoder networks, it then performs two separate segmentation processes depending on the detected defects.
    Finally, using the generated segmentation images, it was able to extract the size distribution and/or location of each defect using the YOLOv5 algorithm for pores and particles, and regression CNNs for GBs.
    
    \added[id=R1, comment={Q2-Q3}]
    {
    We note that the most significant challenge in supervised ML methods for microstructure characterization may be the time-consuming and tedious process required to generate datasets and their corresponding labels for training.
    New technologies and available resources in ML have reduced the need of extensive GPU memory and training time requirements. 
    For instance, Google Colab notebooks may allow the free use of state-of-the-art complex neural networks, such as U-Net and Res-UNet, with up to 12 GBs of GPU within 12 hours of operation.
    However, some ML applications may require very large datasets to achieve accurate predictions, limiting the allowable time and GPU memory from these free sources.
    Additionally, when exportability of a developed ML workflow is required, it may become crucial to explore the reduction of GPU memory and training time from current state-of-the-art networks.
    }
    
    As a result, by leveraging DENSE we optimized the conventional U-Net and Res-UNet for RGB segmentation of grains for a noteworthy reduction of GPU usage of 716 MB, training time reduction of 1 hour, while maintaining high accuracy.
    The DENSE-UNet segmentations tested on multiple materials such as titanium, copper alloy, steel, and Inconel alloy showed accurate RGB segmentations of grain sizes.
    The framework also characterized pores and particles in different materials such as austenite and Ti-6Al-4V, 
    and outputted their size and location from the YOLOv5 algorithm.
        
    While the proposed framework has many advantages, we note certain shortcomings and areas of improvement.
    First, while the framework works for different kinds of microstructures and distributions, a pre-trained network may fall short in performance. 
    When training is required, a minimum number of images are needed to train and outperform conventional algorithms.
    In the case presented in this work, the framework only outperformed the conventional algorithm for datasets larger than 36 images.
    Second, the framework is currently limited to pores, particles, grains and GBs.
    \added[id=R1, comment={Q4}]
    {
        It was demonstrated in this work, that the ML segmentation networks Simple-UNet and DENSE-UNet, and the YOLOv5 object detection network may be capable of characterizing microstructures involving both pores and grain boundaries by using transfer learning. 
        However, in order to develop autonomy for these cases, a larger dataset of microstructures including pores and grains may be required.
        As a future work, the use of larger datasets for training a new extended Classifier CNN to detect these three types of microstructures, may be a promising solution to develop complete and autonomous ML framework. 
    }
    Finally, the framework works on processed images rather than raw experimental data. 
    For instance, SEM data needs to be processed to generate the microstructure images which serve as input for the framework.
    To work with raw data, other ML algorithms must be incorporated into the proposed framework.


\section{Data availability statement}

    {Code available at \url{https://github.com/rperera12/MachineLearning_Framework_Microstructure_Characterization}}


\clearpage

\bibliographystyle{ieeetr}
\bibliography{library}

\begin{thebibliography}{10}

\bibitem{Russell2019Qualification}
R.~Russell, D.~Wells, J.~Waller, B.~Poorganji, E.~Ott, T.~Nakagawa,
  H.~Sandoval, N.~Shamsaei, and M.~Seifi, ``3 - qualification and certification
  of metal additive manufactured hardware for aerospace applications*,'' in
  {\em Additive Manufacturing for the Aerospace Industry} (F.~Froes and
  R.~Boyer, eds.), pp.~33 -- 66, Elsevier, 2019.

\bibitem{Gasser2010Laser}
A.~Gasser, G.~Backes, I.~Kelbassa, A.~Weisheit, and K.~Wissenbach, ``Laser
  additive manufacturing,'' {\em Laser Technik Journal}, vol.~7, no.~2,
  pp.~58--63, 2010.

\bibitem{Su2012Development}
X.~bin SU, Y.~qiang YANG, P.~YU, and J.~feng SUN, ``Development of porous
  medical implant scaffolds via laser additive manufacturing,'' {\em
  Transactions of Nonferrous Metals Society of China}, vol.~22, pp.~s181 --
  s187, 2012.

\bibitem{Woensel2018Printing}
R.~van Woensel, T.~van Oirschot, M.~Burgmans, M.~Mohammadi, and K.~Hermans,
  ``Printing architecture: An overview of existing and promising additive
  manufacturing methods and their application in the building industry,'' {\em
  The International Journal of the Constructed Environment}, vol.~9,
  pp.~57--81, 06 2018.

\bibitem{DeCost2017Characterizing}
B.~L. DeCost and E.~A. Holm, ``Characterizing powder materials using
  keypoint-based computer vision methods,'' {\em Computational Materials
  Science}, vol.~126, pp.~438 -- 445, 2017.

\bibitem{Murr2018Metallographic}
L.~Murr, ``A metallographic review of 3d printing/additive manufacturing of
  metal and alloy products and components,'' {\em Metallography,
  Microstructure, and Analysis}, vol.~7, pp.~103--132, 2018.

\bibitem{Choren2013Youngs}
J.~Choren, S.~Heinrich, and B.~Silver-Thorn, ``Young's modulus and volume
  porosity relationships for additive manufacturing applications,'' {\em
  Journal of Materials Science}, vol.~48, pp.~5103--5112, 08 2013.

\bibitem{FURUMOTO2015Permeability}
T.~Furumoto, A.~Koizumi, M.~R. Alkahari, R.~Anayama, A.~Hosokawa, R.~Tanaka,
  and T.~Ueda, ``Permeability and strength of a porous metal structure
  fabricated by additive manufacturing,'' {\em Journal of Materials Processing
  Technology}, vol.~219, pp.~10 -- 16, 2015.

\bibitem{Tammas-Williams2017FatiguePorosity}
S.~Tammas-Williams, P.~Withers, I.~Todd, and P.~Prangnell, ``The influence of
  porosity on fatigue crack initiation in additively manufactured titanium
  components,'' {\em Scientific Reports}, vol.~7, 08 2017.

\bibitem{SHERIDAN2018Fatigue}
L.~Sheridan, O.~E. Scott-Emuakpor, T.~George, and J.~E. Gockel, ``Relating
  porosity to fatigue failure in additively manufactured alloy 718,'' {\em
  Materials Science and Engineering: A}, vol.~727, pp.~170 -- 176, 2018.

\bibitem{Todaro2020Grain}
C.~Todaro, M.~Easton, D.~Qiu, D.~Zhang, M.~Bermingham, E.~Lui, M.~Brandt,
  D.~Stjohn, and M.~Qian, ``Grain structure control during metal 3d printing by
  high-intensity ultrasound,'' {\em Nature Communications}, vol.~11, p.~142, 01
  2020.

\bibitem{Chapetti2004Fatigue}
M.~Chapetti, H.~Miyata, T.~Tagawa, T.~Miyata, and M.~Fujioka, ``Fatigue
  strength of ultra-fine grained steels,'' {\em Materials Science and
  Engineering: A}, vol.~381, pp.~331--336, 09 2004.

\bibitem{Slotwinski2014Tools}
J.~Slotwinski, E.~Garboczi, P.~Stutzman, C.~Ferraris, S.~Watson, and M.~Peltz,
  ``Characterization of metal powders used for additive manufacturing,'' {\em
  Journal of research of the National Institute of Standards and Technology},
  vol.~119, pp.~460--493, 09 2014.

\bibitem{RABBANI2014An}
A.~Rabbani, S.~Jamshidi, and S.~Salehi, ``An automated simple algorithm for
  realistic pore network extraction from micro-tomography images,'' {\em
  Journal of Petroleum Science and Engineering}, vol.~123, pp.~164 -- 171,
  2014.
\newblock Neural network applications to reservoirs: Physics-based models and
  data models.

\bibitem{LORE2015Facile}
G.~{Lo Re}, F.~Lopresti, G.~Petrucci, and R.~Scaffaro, ``A facile method to
  determine pore size distribution in porous scaffold by using image
  processing,'' {\em Micron}, vol.~76, pp.~37 -- 45, 2015.

\bibitem{Gundersen1985Stereological}
H.~J.~G. Gundersen and E.~B. Jensen, ``Stereological estimation of the
  volume-weighted mean volume of arbitrary particles observed on random
  sections*,'' {\em Journal of Microscopy}, vol.~138, no.~2, pp.~127--142,
  1985.

\bibitem{ASTM2015Standard}
A.~S. for Testing and M.~Philadelphia, {\em {Standard test methods for
  determining average grain size using semiautomatic and automatic image
  analysis}}.
\newblock West Conshohocken, PA: ASTM, 2015.
\newblock Reapproved in 2015.

\bibitem{Qiu2016BigData}
J.~Qiu, Q.~Wu, G.~Ding, Y.~Xu, and S.~Feng, ``A survey of machine learning for
  big data processing,'' {\em EURASIP Journal on Advances in Signal
  Processing}, vol.~2016, 05 2016.

\bibitem{DeCost2016Feedstocks}
B.~DeCost, H.~Jain, A.~Rollett, and E.~Holm, ``Computer vision and machine
  learning for autonomous characterization of am powder feedstocks,'' {\em
  JOM}, vol.~69, 12 2016.

\bibitem{CHOWDHURY2016ImageDriven}
A.~Chowdhury, E.~Kautz, B.~Yener, and D.~Lewis, ``Image driven machine learning
  methods for microstructure recognition,'' {\em Computational Materials
  Science}, vol.~123, pp.~176 -- 187, 2016.

\bibitem{Steinberger2019Dislocations}
D.~Steinberger, H.~Song, and S.~Sandfeld, ``Machine learning-based
  classification of dislocation microstructures,'' {\em Frontiers in
  Materials}, vol.~6, p.~141, 2019.

\bibitem{Furukawa1996Microhardness}
M.~Furukawa, Z.~Horita, M.~Nemoto, R.~Valiev, and T.~Langdon, ``Microhardness
  measurements and the hall-petch relationship in an al-mg alloy with
  submicrometer grain size,'' {\em Acta Materialia}, vol.~44, no.~11, pp.~4619
  -- 4629, 1996.

\bibitem{Tachibana1988Effects}
S.~Tachibana, S.~Kawachi, K.~Yamada, and T.~Kunio, ``Effect of grain refinement
  on the endurance limit of plain carbon steels at various strength levels.,''
  {\em Transactions of the Japan Society of Mechanical Engineers. A}, vol.~54,
  pp.~1956--1961, 1988.

\bibitem{Li2018Automated}
W.~Li, K.~G. Field, and D.~Morgan, ``Automated defect analysis in electron
  microscopic images,'' {\em npj Computational Materials}, vol.~4, p.~36, Jul
  2018.

\bibitem{ANDERSON2020Automated}
C.~M. Anderson, J.~Klein, H.~Rajakumar, C.~D. Judge, and L.~K. Béland,
  ``Automated detection of helium bubbles in irradiated x-750,'' {\em
  Ultramicroscopy}, vol.~217, p.~113068, 2020.

\bibitem{Li2020supportvector}
M.~Li, D.~Chen, S.~Liu, and D.~Guo, ``Online learning method based on support
  vector machine for metallographic image segmentation,'' {\em Signal, Image
  and Video Processing}, 09 2020.

\bibitem{BASKARAN2020Adaptive}
A.~Baskaran, G.~Kane, K.~Biggs, R.~Hull, and D.~Lewis, ``Adaptive
  characterization of microstructure dataset using a two stage machine learning
  approach,'' {\em Computational Materials Science}, vol.~177, p.~109593, 2020.

\bibitem{Jang2020Residual}
J.~Jang, D.~Van, H.~Jang, D.~H. Baik, S.~D. Yoo, J.~Park, S.~Mhin, J.~Mazumder,
  and S.~H. Lee, ``Residual neural network-based fully convolutional network
  for microstructure segmentation,'' {\em Science and Technology of Welding and
  Joining}, vol.~25, no.~4, pp.~282--289, 2020.

\bibitem{Roberts2019DefectSegNet}
G.~Roberts, S.~Y. Haile, R.~Sainju, D.~J. Edwards, B.~Hutchinson, and Y.~Zhu,
  ``Deep learning for semantic segmentation of defects in advanced stem images
  of steels,'' {\em Scientific Reports}, vol.~9, p.~12744, Sep 2019.

\bibitem{Kasim2020Building}
M.~F. Kasim, D.~Watson-Parris, L.~Deaconu, S.~Oliver, P.~Hatfield, D.~H.
  Froula, G.~Gregori, M.~Jarvis, S.~Khatiwala, J.~Korenaga,
  J.~Topp-Mugglestone, E.~Viezzer, and S.~M. Vinko, ``Building high accuracy
  emulators for scientific simulations with deep neural architecture search,''
  2020.

\bibitem{Lehto2014Influence}
P.~Lehto, H.~Remes, T.~Saukkonen, H.~Hänninen, and J.~Romanoff, ``Influence of
  grain size distribution on the hall–petch relationship of welded structural
  steel,'' {\em Materials Science and Engineering: A}, vol.~592, pp.~28 -- 39,
  2014.

\bibitem{Lehto2016Characterization}
P.~Lehto, H.~Remes, T.~Sarikka, and J.~Romanoff, ``Characterisation of local
  grain size variation of welded structural steel,'' {\em Welding in the
  World}, vol.~60, pp.~673 -- 688, 2016.

\bibitem{Al-Wzwazy2016Handwritten}
H.~Al-Wzwazy, ``Handwritten digit recognition using convolutional neural
  networks,'' {\em International Journal of Innovative Research in Computer and
  Communication Engineering}, vol.~4, 02 2016.

\bibitem{Shang2016Spam}
E.~{Shang} and H.~{Zhang}, ``Image spam classification based on convolutional
  neural network,'' in {\em 2016 International Conference on Machine Learning
  and Cybernetics (ICMLC)}, vol.~1, pp.~398--403, 2016.

\bibitem{Bhatnagar2017Fashion}
S.~{Bhatnagar}, D.~{Ghosal}, and M.~H. {Kolekar}, ``Classification of fashion
  article images using convolutional neural networks,'' in {\em 2017 Fourth
  International Conference on Image Information Processing (ICIIP)}, pp.~1--6,
  2017.

\bibitem{Levi2015Age}
G.~Levi and T.~Hassner, ``Age and gender classification using convolutional
  neural networks,'' in {\em Proceedings of the IEEE Conference on Computer
  Vision and Pattern Recognition (CVPR) Workshops}, June 2015.

\bibitem{Sharma2018RealTime}
N.~Sharma, V.~Jain, and A.~Mishra, ``An analysis of convolutional neural
  networks for image classification,'' {\em Procedia Computer Science},
  vol.~132, pp.~377 -- 384, 2018.
\newblock International Conference on Computational Intelligence and Data
  Science.

\bibitem{Kaibo2003SoftMax}
K.~Duan, S.~S. Keerthi, W.~Chu, S.~K. Shevade, and A.~N. Poo, ``Multi-category
  classification by soft-max combination of binary classifiers,'' in {\em
  Multiple Classifier Systems} (T.~Windeatt and F.~Roli, eds.), (Berlin,
  Heidelberg), pp.~125--134, Springer Berlin Heidelberg, 2003.

\bibitem{Xia2019LandmarkPose}
J.~{Xia}, L.~{Cao}, G.~{Zhang}, and J.~{Liao}, ``Head pose estimation in the
  wild assisted by facial landmarks based on convolutional neural networks,''
  {\em IEEE Access}, vol.~7, pp.~48470--48483, 2019.

\bibitem{Chenjing2014AgeCNN}
C.~Yan, C.~Lang, T.~Wang, X.~Du, and C.~Zhang, ``Age estimation based on
  convolutional neural network,'' in {\em Advances in Multimedia Information
  Processing -- PCM 2014} (W.~T. Ooi, C.~G.~M. Snoek, H.~K. Tan, C.-K. Ho,
  B.~Huet, and C.-W. Ngo, eds.), (Cham), pp.~211--220, Springer International
  Publishing, 2014.

\bibitem{Chen2018StockMarket}
S.~Chen and H.~He, ``Stock prediction using convolutional neural network,''
  {\em {IOP} Conference Series: Materials Science and Engineering}, vol.~435,
  p.~012026, nov 2018.

\bibitem{Cipolla2017SegNet}
V.~{Badrinarayanan}, A.~{Kendall}, and R.~{Cipolla}, ``Segnet: A deep
  convolutional encoder-decoder architecture for image segmentation,'' {\em
  IEEE Transactions on Pattern Analysis and Machine Intelligence}, vol.~39,
  no.~12, pp.~2481--2495, 2017.

\bibitem{Ronneberger2015unet}
O.~Ronneberger, P.~Fischer, and T.~Brox, ``U-net: Convolutional networks for
  biomedical image segmentation,'' {\em CoRR}, vol.~abs/1505.04597, 2015.

\bibitem{He2015Residual}
K.~He, X.~Zhang, S.~Ren, and J.~Sun, ``Deep residual learning for image
  recognition,'' {\em CoRR}, vol.~abs/1512.03385, 2015.

\bibitem{Philipp2017Gradients}
G.~Philipp, D.~Song, and J.~G. Carbonell, ``Gradients explode - deep networks
  are shallow - resnet explained,'' 2018.

\bibitem{Xiao2018Retina}
X.~{Xiao}, S.~{Lian}, Z.~{Luo}, and S.~{Li}, ``Weighted res-unet for
  high-quality retina vessel segmentation,'' in {\em 2018 9th International
  Conference on Information Technology in Medicine and Education (ITME)},
  pp.~327--331, 2018.

\bibitem{Feng2020Photoacoustic}
J.~Feng, J.~Deng, Z.~Li, Z.~Sun, H.~Dou, and K.~Jia, ``End-to-end res-unet
  based reconstruction algorithm for photoacoustic imaging,'' {\em Biomed. Opt.
  Express}, vol.~11, pp.~5321--5340, Sep 2020.

\bibitem{Zhang2019Photovoltaic}
H.~Zhang, X.~Hong, S.~Zhou, and Q.~Wang, ``Infrared image segmentation for
  photovoltaic panels based on res-unet,'' in {\em Pattern Recognition and
  Computer Vision} (Z.~Lin, L.~Wang, J.~Yang, G.~Shi, T.~Tan, N.~Zheng,
  X.~Chen, and Y.~Zhang, eds.), (Cham), pp.~611--622, Springer International
  Publishing, 2019.

\bibitem{Hai2019BrainTumor}
H.~Xu, H.~Xie, Y.~Liu, C.~Cheng, C.~Niu, and Y.~Zhang, ``Deep cascaded
  attention network for multi-task brain tumor segmentation,'' in {\em Medical
  Image Computing and Computer Assisted Intervention -- MICCAI 2019} (D.~Shen,
  T.~Liu, T.~M. Peters, L.~H. Staib, C.~Essert, S.~Zhou, P.-T. Yap, and
  A.~Khan, eds.), (Cham), pp.~420--428, Springer International Publishing,
  2019.

\bibitem{Hansen2016CMA}
N.~Hansen, ``The {CMA} evolution strategy: {A} tutorial,'' {\em CoRR},
  vol.~abs/1604.00772, 2016.

\bibitem{Girshick2014Rich}
R.~Girshick, J.~Donahue, T.~Darrell, and J.~Malik, ``Rich feature hierarchies
  for accurate object detection and semantic segmentation,'' 2014.

\bibitem{Redmon2015You}
J.~Redmon, S.~K. Divvala, R.~B. Girshick, and A.~Farhadi, ``You only look once:
  Unified, real-time object detection,'' {\em CoRR}, vol.~abs/1506.02640, 2015.

\bibitem{Redmon2016YOLO9000}
J.~Redmon and A.~Farhadi, ``{YOLO9000:} better, faster, stronger,'' {\em CoRR},
  vol.~abs/1612.08242, 2016.

\bibitem{Redmon2018YOLOv3}
J.~Redmon and A.~Farhadi, ``Yolov3: An incremental improvement,'' {\em CoRR},
  vol.~abs/1804.02767, 2018.

\bibitem{Bochkovskiy2020YOLOv4}
A.~Bochkovskiy, C.-Y. Wang, and H.-Y.~M. Liao, ``Yolov4: Optimal speed and
  accuracy of object detection,'' {\em ArXiv}, 2020.

\bibitem{Jocher2020yolov5}
G.~Jocher, A.~Stoken, J.~Borovec, NanoCode012, ChristopherSTAN, L.~Changyu,
  Laughing, tkianai, A.~Hogan, lorenzomammana, yxNONG, AlexWang1900,
  L.~Diaconu, Marc, wanghaoyang0106, ml5ah, Doug, F.~Ingham, Frederik, Guilhen,
  Hatovix, J.~Poznanski, J.~Fang, L.~Yu, 
  changyu98, M.~Wang, N.~Gupta, O.~Akhtar, PetrDvoracek, and P.~Rai,
  ``{ultralytics/yolov5: v3.1 - Bug Fixes and Performance Improvements},'' {\em
  Towards AI - Multidisciplinary Science Journal}, Oct. 2020.

\bibitem{Bisong2019Google}
E.~Bisong, {\em Google Colaboratory}, pp.~59--64.
\newblock Berkeley, CA: Apress, 2019.

\bibitem{NEURIPS2019PyTorch}
A.~Paszke, S.~Gross, F.~Massa, A.~Lerer, J.~Bradbury, G.~Chanan, T.~Killeen,
  Z.~Lin, N.~Gimelshein, L.~Antiga, A.~Desmaison, A.~Kopf, E.~Yang, Z.~DeVito,
  M.~Raison, A.~Tejani, S.~Chilamkurthy, B.~Steiner, L.~Fang, J.~Bai, and
  S.~Chintala, ``Pytorch: An imperative style, high-performance deep learning
  library,'' in {\em Advances in Neural Information Processing Systems 32}
  (H.~Wallach, H.~Larochelle, A.~Beygelzimer, F.~d\textquotesingle
  Alch\'{e}-Buc, E.~Fox, and R.~Garnett, eds.), pp.~8024--8035, Curran
  Associates, Inc., 2019.

\bibitem{Paszke2017Automatic}
A.~Paszke, S.~Gross, S.~Chintala, G.~Chanan, E.~Yang, Z.~DeVito, Z.~Lin,
  A.~Desmaison, L.~Antiga, and A.~Lerer, ``Automatic differentiation in
  pytorch,'' {\em NIPS}, 2017.

\bibitem{Holm2016Synthetic}
B.~L. DeCost and E.~A. Holm, ``A large dataset of synthetic sem images of
  powder materials and their ground truth 3d structures,'' {\em Data in Brief},
  vol.~9, pp.~727 -- 731, 2016.

\bibitem{Li2018Selective}
X.~Li, H.~J. Willy, S.~Chang, W.~Lu, T.~S. Herng, and J.~Ding, ``Selective
  laser melting of stainless steel and alumina composite: Experimental and
  simulation studies on processing parameters, microstructure and mechanical
  properties,'' {\em Materials and Design}, vol.~145, pp.~1 -- 10, 2018.

\bibitem{Kalra2016Canny}
A.~{Kalra} and R.~L. {Chhokar}, ``A hybrid approach using sobel and canny
  operator for digital image edge detection,'' in {\em 2016 International
  Conference on Micro-Electronics and Telecommunication Engineering (ICMETE)},
  pp.~305--310, 2016.

\end{thebibliography}
\end{document}